\documentstyle[11pt]{article}
\renewcommand{\ll}{\lambda_n}
        \newcommand{\mm}{\mu}

\newcommand{\hPPo}{e^{\psi(\lambda)+\tau(\lambda)}e^{-\frac12(\psi(\mm)+
\tau(\mm))} }

\newcommand{\be}{\begin{equation}}
\newcommand{\ee}{\end{equation}}
\newcommand{\ba}{\begin{eqnarray}}
\newcommand{\ea}{\end{eqnarray}}
\newcommand{\non}{\nonumber\\}

\newcommand{\Eq}[1]{(\ref{#1})}
\newcommand{\one}{\vspace{-0.1em}}                                  
\renewcommand{\Im}{\mathop{\rm Im}\nolimits}                        
\newcommand{\tr}{\mathop{\rm tr}\nolimits}
\newcommand{\dis}{\displaystyle}
\newcommand{\hr}{\hat \rho}
\newcommand{\llb}{{\lambda_{N_2}}}               
\newcommand{\lls}{{\lambda_{N_1}}}

\newcommand{\cMi}[1]{{\cal M}{\it i}\left\{#1\right\}}
\newcommand{\cM}{{\cal M}{\it i}}
\newcommand{\Co}{{\cal O}(1/L^2)}
\newcommand{\co}{{\cal O}(1/L)}

\newcommand{\hS}{{\widehat S}}

\newcommand{\hrp}{{2\pi\hat \rho}}
\newcommand{\hU}{{\widehat U}}
\newcommand{\hQ}{{\widehat Q}}

\newcommand{\ket}{{|0\rangle}}
\newcommand{\bra}{{\langle 0|}}
\newcommand{\ketd}{{|0)}}
\newcommand{\brad}{{(0|\,}}

\newcommand{\hxi}{\hat \xi}
\newcommand{\hxx}{\hat \xi(\mu)}
\newcommand{\hxxj}{\hat\xi(\mu_j)}          
\newcommand{\hxxk}{\hat\xi(\mu_k)}          

        \newcommand{\lm}{\ll-\mm}

        \newcommand{\stint}{\int\limits_{-\infty}^\infty}
\newcommand{\qint}{\int\limits_{-q}^q}
        \newcommand{\mint}{\int\limits_\lls^\llb
\hspace{-0.43cm}\backslash}
        \newcommand{\pint}{\int\limits_{-\infty}^\infty
\hspace{-0.43cm}\backslash}

\newlength{\minitwocolumn}
\catcode`\@=11
\def\relaxnext@{\let\next\relax}

\def\eq#1\endeq{\begin{eqnarray}#1\end{eqnarray}}
\def\eqn#1\endeqn{\begin{eqnarray*}#1\end{eqnarray*}}

\makeatletter
\@addtoreset{equation}{section}
\makeatother

\newtheorem{thm}{Theorem}[section]

\newtheorem{lem}[thm]{Lemma}

\begin{document}
\begin{flushright}
January 1997\\
ITP-SUNY-SB-96-71
\end{flushright}
\vspace{24pt}
\begin{center}
\begin{Large}
{\bf Determinant representation for dynamical correlation
functions of the  Quantum nonlinear Schr\"odinger equation.}

\end{Large}

~

(Short title : Determinant representation for correlation functions
of Bose gas)

\vspace{26pt}
T.~Kojima\raisebox{2mm}{{\scriptsize a}$\star$},
V.~E.~Korepin\raisebox{2mm}{{\scriptsize b}}

and N.~A.~Slavnov\raisebox{2mm}{{\scriptsize c}}

\vspace{6pt}

~\raisebox{2mm}{{\scriptsize a}}
{\it Research Institute for Mathematical Sciences,
     Kyoto University, Kyoto 606, Japan}

~\raisebox{2mm}{{\scriptsize b}}
{\it Institute for Theoretical Physics, State University of 
New York at Stony Brook,
Stony Brook, NY 11794-3840, U. S. A.}

~\raisebox{2mm}{{\scriptsize c}}
{\it Steklov Mathematical Institute,
Gubkina 8, Moscow 117966, Russia.}
\vspace{42pt}

\underline{Abstract}
\end{center}
Painlev\'e analysis of correlation functions of the impenetrable Bose gas
by M.~Jimbo, T.~Miwa, Y.~Mori and M.~Sato \cite{J.M.M.S.} was based on the
determinant representation of these correlation functions obtained
by A.~Lenard \cite{Len}. The impenetrable Bose gas is the free
fermionic case of
the quantum nonlinear Schr\"odinger equation.
In this paper we generalize the Lenard determinant representation
for $\langle\psi(0,0)\psi^\dagger(x,t)\rangle$ to the non-free fermionic case.
We also include time and temperature dependence. In
forthcoming publications we shall perform
 the JMMS analysis of this correlation function. This will give
us a completely integrable equation and asymptotic for
the quantum correlation function of interacting fermions.
\vspace{14pt}

\vfill
\hrule

\vskip 3mm
\begin{footnotesize}
\noindent
\raisebox{2mm}{$a$}
kojima@kurims.kyoto-u.ac.jp,~
\raisebox{2mm}{$b$}
korepin@insti.physics.sunysb.edu,~
\raisebox{2mm}{$c$}
nslavnov@class.mi.ras.ru\\
\noindent\raisebox{2mm}{$\star$}
Research Fellow of the Japan Society
for the Promotion of Science.
\end{footnotesize}
\newpage

\section{Introduction}
We consider  exactly solvable models of statistical mechanics in one space
and one time dimension. The Quantum Inverse Scattering Method
and Algebraic Bethe Ansatz are effective methods for a description of the
spectrum of these models. Our aim is the evaluation of correlation functions of
exactly solvable models. Our approach is based on the determinant representation for correlation functions. It consists of a few steps: first the correlation
function is represented as a determinant of a Fredholm integral operator,
second --- the Fredholm integral operator is described by a 
classical completely
integrable equation, third --- the classical completely integrable equation
is solved by means of the Riemann-Hilbert problem. This permits us to evaluate
the long distance and large time asymptotics of the correlation function.
The method is described in \cite{J.M.M.S.}, \cite{K.B.I.}.

The most interesting correlation functions
are time dependent correlation functions. The determinant
representation for  time dependent correlation functions was known only
for the impenetrable Bose gas (the spectrum of the Hamiltonian of this model
is equivalent to free fermions). In this paper we have found the determinant 
representation for the time dependent correlation function of local fields
of the penetrable Bose gas. The main idea for the construction of the determinant representation is the following. We introduce auxiliary Bose fields (acting in the
canonical Fock space) in order to remove the two body scattering matrix and to
reduce the model to the free fermionic case. We want to emphasize that all
dual fields, which we introduce commute (belong to the same Abelian
sub-algebra). Therefore we do not have any ordering problem. This will
also permit us to perform nonperturbative calculations, which are necessary for
the derivation of the integrable equation for the correlation function.

First we shall discuss our model. 

Quantum nonlinear Schr\"odinger equation (equivalent to
 Bose gas with delta-function interaction) can be
 described by the canonical Bose fields
$\psi(x)$ and $\psi^{\dagger}(x)$ with the commutation relations:
\begin{eqnarray}
[\psi(x), \psi^{\dagger}(y)]=\delta(x-y),~~
[\psi(x), \psi(y)]=[\psi^{\dagger}(x), \psi^{\dagger}(y)]=0,
\end{eqnarray}
acting in the Fock space. Fock vacuum $|0\rangle$ and dual vector
$\langle0|$ are important. They are defined by the relations
\begin{equation}
\psi(x) \vert 0 \rangle =0,\qquad
\langle 0 \vert \psi^{\dagger}(x)=0,\qquad \langle 0 \mid 0 \rangle=1.
\end{equation}

The Hamiltonian of the model is
\begin{eqnarray}
{H}=\int dx
\left({\partial_x}\psi^{\dagger}(x)
{\partial_x} \psi(x)+
c\psi^{\dagger}(x)\psi^{\dagger}(x)\psi(x)\psi(x)
-h \psi^{\dagger}(x)\psi(x)\right),
\end{eqnarray}
Here $c$ is the coupling constant and $h>0$
is the chemical potential. We shall consider the repulsive case
$0<c\le\infty$.

The spectrum of the model was first described by E.~H.~Lieb and
 W.~Liniger  \cite{L}, \cite{L.L.}.
The Lax representation for the corresponding classical equation of 
 motion
\begin{eqnarray}
i\frac{\partial}{\partial t} \psi=
[\psi, H]=
-\frac{\partial^2}{{\partial x}^2} \psi
+2c \psi^{\dagger}\psi \psi
-h \psi,
\end{eqnarray}
was found by V.~E.~Zakharov and A.~B.~Shabat \cite{ZS}. 
The Quantum Inverse Scattering Method for the model
was formulated by L.~D.~Faddeev and E.~K.~Sklyanin 
\cite{FS}.

In this paper we shall follow the notations of  \cite{K.B.I.}.
First the model
is considered in a finite periodic box of  length $L$. Later the
thermodynamic limit is considered when the length of the box $L$ and
the number of particles in the ground state go to infinity, with the 
ratio $N/L$ held fixed.

The Quantum nonlinear Schr\"odinger equation is equivalent to the Bose gas
with delta-function interaction. In the sector with $N$ particles the
Hamiltonian of Bose gas is given by
\begin{eqnarray}
{\cal H}_N=-\sum_{j=1}^N
\frac{\partial ^2}{\partial z_j^2}
+2c \sum_{1 \leq j < k \leq N}
\delta(z_k-z_j)-Nh.
\end{eqnarray}
Now a few words about the 
organization of the paper.

In Section~2 we shall review the Algebraic Bethe Ansatz and collect all the known facts necessary for further calculations. In Section~3 we shall calculate 
the form factor of the local field in  finite volume. 
In Section~4 we shall present the idea of summation with respect 
to all intermediate states. In Section~5 we introduce an auxiliary Bosonic Fock space and auxiliary Bose fields. This helps us to represent the correlation
function as a determinant in the finite volume. In Section~6 we consider
the thermodynamic limit of the determinant representation for correlation function.
Length of the periodic box $L$ and number of particles in the ground state
go to infinity but their ratio remains fixed. This leads us to the main result
of the paper (see formul\ae~\Eq{corr10}--\Eq{proect1}). The 
 correlation function of local fields in the infinite volume
is represented as a determinant of a Fredholm integral operator. For
evaluation of the thermodynamic limit it is necessary to sum up singular 
expressions. Appendix~A is devoted to these summations.
In Appendix~B we present realization of quantum dual fields 
as linear combinations of the
canonical Bose fields. Appendix C shows how to reduce  the number of dual fields.
Appendix D contains determinant representation for temperature correlation
function.

In forthcoming publications 
we shall use the determinant representation for the derivation
of completely integrable equation for correlation functions. Later we
shall solve this equation by means of the Riemann-Hilbert problem and
evaluate the long-distance asymptotic.

\section{Algebraic Bethe Ansatz}
Let us review some main features of the Algebraic
 Bethe Ansatz, which we shall use later.
We consider the quantum nonlinear 
Schr\"odinger model.
The starting point and central object of
the Quantum Inverse Scattering Method is the R-matrix,
which is a solution of the Yang-Baxter equation.
For the case of the quantum nonlinear Schr\"odinger equation,
it is of the form :
\begin{eqnarray}
R(\lambda,\mu)=\left(
\begin{array}{cccc}
f(\mu,\lambda)&0&0&0\\
0&g(\mu,\lambda)&1&0\\
0&1&g(\mu,\lambda)&0\\
0&0&0&f(\mu,\lambda)
\end{array}
\right),
\end{eqnarray}
where
\begin{eqnarray}
g(\lambda,\mu)=\frac{ic}{\lambda-\mu},~~~
f(\lambda,\mu)=\frac{\lambda-\mu+ic}{\lambda-\mu}.
\end{eqnarray}
Later we shall also use functions
\begin{eqnarray}
h(\lambda,\mu)=\frac{\lambda-\mu+ic}{ic},~~~
t(\lambda,\mu)=\frac{(ic)^2}{(\lambda-\mu)(\lambda-\mu+ic)}
=\frac{g(\lambda,\mu)}{h(\lambda,\mu)}.
\end{eqnarray}

Another important object is the monodromy matrix
\begin{equation}
T(\lambda)=\left(
\begin{array}{cc}
A(\lambda)&B(\lambda)\\
C(\lambda)&D(\lambda)
\end{array}
\right)
\end{equation}
The operators $A,~B,~C,~D$ are acting in the  Fock space
where the operator $\psi(x)$ was defined. Their commutation
relations are given by
\begin{eqnarray}
R(\lambda,\mu)\left(T(\lambda)\otimes T(\mu)\right)
=\left(T(\mu)\otimes T(\lambda)\right)R(\lambda,\mu).
\end{eqnarray}
These relations are written out explicitly  in Section VII.1 of \cite{K.B.I.}.

The hermiticity properties of $T(\lambda)$ are
\begin{equation}
\sigma_xT^*(\bar\lambda)\sigma_x=T(\lambda),
\end{equation}
so that $B^\dagger(\lambda)=C(\bar\lambda)$.

The Hamiltonian of the model can be expressed in terms of
$A(\lambda)+D(\lambda)$ by means of trace identities (Section
VI.3 of \cite{K.B.I.}).
The vacuum is eigenvector of the diagonal elements of $T(\lambda)$
\begin{eqnarray}
A(\lambda)|0\rangle=a(\lambda)|0\rangle; &\qquad &
D(\lambda)|0\rangle=d(\lambda)|0\rangle\\
\langle0|A(\lambda)=a(\lambda)\langle0|; &\qquad &
\langle0|D(\lambda)=d(\lambda)\langle0|;\\
a(\lambda)=\exp\left\{-\frac{iL\lambda}2\right\};&\qquad &
d(\lambda)=\exp\left\{\frac{iL\lambda}2\right\}
\end{eqnarray}
Later we shall also use the function
\begin{equation}
r(\lambda)=\frac{a(\lambda)}{d(\lambda)}=e^{-i\lambda L}
\end{equation}
The operator $C(\lambda)$ annihilates the vacuum vector and the operator $B(\lambda)$ 
annihilates the dual vacuum:
\begin{equation}
C(\lambda)|0\rangle=0,\qquad
\langle0|B(\lambda)=0.
\end{equation}

The Hamiltonian of the model commutes with $A(\lambda)+D(\lambda)$ and they
can be diagonalized simultaneously. The eigenvectors of the Hamiltonian are
\begin{equation}
\prod_{j=1}^NB(\mu_j)|0\rangle,\qquad\mbox{and}\qquad
\langle0|\prod_{j=1}^NC(\mu_j),
\end{equation}
if ${\mu_j}$ satisfy Bethe Equations
\begin{equation}\label{BETHEeq}
\frac{a(\mu_j)}{d(\mu_j)}\prod_{k=1\atop{k\ne j}}^N
\frac{f(\mu_j,\mu_k)}{f(\mu_k,\mu_j)}=1,\qquad\mbox{or}\qquad
\frac{a(\mu_j)}{d(\mu_j)}\prod_{k=1}^N
\frac{h(\mu_j,\mu_k)}{h(\mu_k,\mu_j)}=(-1)^{N-1}
\end{equation}
It is convenient to rewrite \Eq{BETHEeq} in logarithmic form. For the
ground state 
\begin{eqnarray}\label{phi}
\varphi_j+\pi\equiv L\mu_j+\sum_{k=1}^N
i\ln\left(\frac{ic+\mu_j-\mu_k}{ic-\mu_j+\mu_k}\right)=2\pi
\left(j-\frac{N+1}2\right)
\end{eqnarray}
 It is proven in Section I.2 of \cite{K.B.I.} that
solutions $\mu_j$  of equation \Eq{phi} 
are real.

The distribution of $\mu_j$ in the ground state in thermodynamic limit can be described by linear integral equation. The thermodynamic limit is defined in the following way: $N\to\infty,\quad L\to\infty$ and $N/L=D$ is fixed. In this limit
$\mu_j$ condense ($\mu_{j+1}-\mu_j={\cal O}(1/L)$) and fill the symmetric interval $[-q,q]$,
 where $q$ is  the value of spectral parameter on the Fermi surface.

 In the thermodynamic limit the function of local density
$\rho(\mu)$ can be defined in the following way
\begin{equation}
\rho(\mu_j)=\lim\frac1{L(\mu_{j+1}-\mu_j)}.
\end{equation}
The $\lim$ in the r.h.s. denotes the thermodynamic limit.
This function satisfies the Lieb-Liniger integral equation
\be\label{density}
\rho(\mu) -\frac{1}{2\pi}\int\limits_{-q}^q
K(\nu,\mu)\rho(\nu)d\nu = \frac{1}{2\pi}.
\ee
Here
\be
K(\nu,\mu)=\frac{2c}{c^2+(\mu-\nu)^2},
\ee
and
\be
D=\frac NL=\qint\,d\mu\rho(\mu).
\ee
In such a way we have described the ground state.

Now we can define the correlation function of the local fields
\be\label{defcor}
\langle \psi(0,0)\psi^\dagger(x,t)\rangle
=\lim
\frac{\dis
\bra\prod_{j=1}^N C(\mu_j)\psi(0,0)
\psi^\dagger(x,t)\prod_{j=1}^{N}
B(\mu_j)\ket
}{\dis
\bra\prod_{j=1}^{N} C(\mu_j)\prod_{j=1}^{N}
B(\mu_j)\ket
}.
\ee
Here
\be
\psi^\dagger(x,t)=e^{iHt}\psi^\dagger(x,0)e^{-iHt}.
\ee
We shall use the notation $\mu_j$ for the ground state only. The square of the
norm of the ground state wave function (denominator of the 
correlation function)
was found in \cite{KOR},
\ba\label{normBETHE}
\bra\prod_{j=1}^NC(\mu_j)\prod_{j=1}^{N}B(\mu_j)\ket
&=&
c^N\left( \prod_{N\geq j>k\geq 1}g(\mu_j,\mu_k)g(\mu_k,\mu_j)\right)
\non
&&\non
&&\times\left(\prod_{j=1}^N\prod_{k=1}^N h(\mu_j,\mu_k)\right)
{\det}_N\frac{\partial \varphi_j}{\partial \mu_k}.
\ea
Here $\partial\varphi_j/\partial\mu_k$ is $N\times N$ matrix
\be
\frac{\partial\varphi_j}{\partial\mu_k}=
\delta_{jk}\left[L+\sum_{l=1}^NK(\mu_j,\mu_l)\right]
-K(\mu_j,\mu_k).
\ee
Let us emphasize that $\det(\partial\varphi_j/\partial\mu_k)>0$
(see Section I.2 of \cite{K.B.I.}). 
The thermodynamic limit of the square of the norm can be described by the 
following formula:
\be\label{normlimit}
\lim \left(\frac{\dis
{\det}_N\frac{\partial \varphi_j}{\partial \mu_k}}
{\prod_{j=1}^N 2\pi L \rho(\mu_j)}\right)=
\det\left(\hat I-\frac{1}{2\pi}\hat K\right),
\ee
where $\hat K$ is an integral operator acting on some trial function
$f(\lambda)$ as
\be
(\hat Kf)(\lambda)=\qint K(\lambda,\mu)f(\mu)\,d\mu.
\ee
The proof can be found in \cite{KOR} (see also Section 
X.4 of \cite{K.B.I.}).

In order to calculate the correlation function we shall also need a description
of excited states. We need to consider excited states which have one more 
particle than in the ground state
\begin{equation}
\prod_{j=1}^{N+1}B(\lambda_j)|0\rangle,\qquad\mbox{and}\qquad
\langle0|\prod_{j=1}^{N+1}C(\lambda_j),
\end{equation}
where ${\lambda_j}$ have to satisfy Bethe Equations
\begin{equation}\label{BETheeq}
\frac{a(\lambda_j)}{d(\lambda_j)}\prod_{k=1\atop{k\ne j}}^{N+1}
\frac{f(\lambda_j,\lambda_k)}{f(\lambda_k,\lambda_j)}=1,\qquad\mbox{or}\qquad
\frac{a(\lambda_j)}{d(\lambda_j)}\prod_{k=1}^{N+1}
\frac{h(\lambda_j,\lambda_k)}{h(\lambda_k,\lambda_j)}=(-1)^{N}
\end{equation}
We shall further assume that the  number of particles in the
ground state $N$ is even. In order to write the logarithmic form of the Bethe
Equations it is convenient to introduce
\begin{equation}\label{tildephi}
\tilde\varphi_j\equiv
 L\lambda_j+\sum_{k=1\atop{k\ne j}}^{N+1}
i\ln\left(\frac{\lambda_j-\lambda_k+ic}{\lambda_j-\lambda_k-ic}\right).
\end{equation}
The  Bethe equations can now be written as
\be\label{tildevarphi}
\tilde\varphi_j=2\pi n_j,
\ee
where $n_j$ is an ordered set of different integer numbers
$n_{j+1}>n_j$. One can prove  that
all $\lambda_j$ are real. In order to enumerate all the eigenstates in the sector with $N+1$ particles we have to consider all sets of ordered integers $n_j$.
The square of the norm of the excited state is
\ba
\bra\prod_{j=1}^{N+1}C(\lambda_j)\prod_{j=1}^{N+1}B(\lambda_j)\ket
&=&
c^{N+1}\left( \prod_{N+1\geq j>k\geq 1}g(\lambda_j,\lambda_k)
g(\lambda_k,\lambda_j)\right)
\non
&&\non
&&\times\left(\prod_{j=1}^{N+1}\prod_{k=1}^{N+1} h(\lambda_j,\lambda_k)\right)
{\det}_{N+1}\frac{\partial\tilde \varphi_j}{\partial \lambda_k}.
\ea

For the excited state $\det(\partial\tilde\varphi_j/\partial\lambda_k)$
is also positive.
We shall also mention that the scattering matrix of elementary excitations can be found in Section I.4 of \cite{K.B.I.}. It depends strongly on momenta, this
shows that the model is not free fermionic.

Now we can define the form factor in the finite volume
\be
F_N =
\bra\prod_{j=1}^N C(\mu_j)\psi(0,0)\prod_{j=1}^{N+1}B(\lambda_j)\ket.
\label{FormFactor}
\ee
We shall calculate it in the next section. We shall also need the conjugated 
form factor
\be
\begin{array}{l}
{\dis
\bra\prod_{j=1}^{N+1} C(\lambda_j)\psi^\dagger(x,t)\prod_{j=1}^{N}
B(\mu_j)\ket}\\
\\
\qquad={\dis
e^{-iht}\cdot \exp\left[it\left(\sum_{j=1}^{N+1}\lambda^2_j-\sum_{k=1}^N
\mu_k^2\right)
 -ix\left(\sum_{j=1}^{N+1}\lambda_j-\sum_{k=1}^N
\mu_k\right)\right]\cdot\overline{F}_N}.
\end{array}
\label{cc1}
\ee
Here we used the fact that the energy and momentum of the eigenstate are given
by the expressions
\be
E_{N+1}=\sum_{j=1}^{N+1}(\lambda_j^2-h),
\ee
\be
P_{N+1}=\sum_{j=1}^{N+1}\lambda_j.
\ee

\section{Form Factor}
The main purpose of the paper is to evaluate the correlation function.
In the finite volume we shall use the  notation
\be
\langle \psi(0,0)\psi^\dagger(x,t) \rangle_N
=\frac{\bra \prod_{j=1}^NC(\mu_j) 
\psi(0,0)\psi^\dagger(x,t)
\prod_{j=1}^NB(\mu_j)\ket}
{\bra\prod_{j=1}^NC(\mu_j)\prod_{j=1}^NB(\mu_j)\ket}.
\ee

We shall use the standard representation of correlation function in terms
of the form factors
\be
\begin{array}{l}
\langle \psi(0,0)\psi^\dagger(x,t)\rangle_N\\
={\dis \sum_{\mbox{all }\{\lambda\}_{N+1}}}
\frac{\dis
\bra\prod_{j=1}^N C(\mu_j)\psi(0,0)\prod_{j=1}^{N+1}B(\lambda_j)\ket
\bra\prod_{j=1}^{N+1} C(\lambda_j)\psi^\dagger(x,t)\prod_{j=1}^{N}
B(\mu_j)\ket
}{\dis
\bra\prod_{j=1}^{N+1} C(\lambda_j)\prod_{j=1}^{N+1}B(\lambda_j)\ket
\bra\prod_{j=1}^{N} C(\mu_j)\prod_{j=1}^{N}
B(\mu_j)\ket
}\end{array}.
\ee

In order to calculate the form factor we need to know the action of the local field on the eigenvector. This can be found in \cite{IKR} (see also
Section XII.2 of \cite{K.B.I.}).
\be
\psi(0,0)\prod_{j=1}^{N+1}B(\lambda_j)\ket
=
-i\sqrt{c}\sum_{\ell=1}^{N+1}a(\lambda_\ell)
\left(\prod_{m=1\atop{m\ne \ell}}^{N+1}f(\lambda_\ell, \lambda_m)\right)
\prod_{m=1\atop{m\ne\ell}}^{N+1}B(\lambda_m)\ket,
\ee
This permits us to represent form factor as follows
\ba
F_N&=& -i\sqrt{c}\sum_{\ell=1}^{N+1}a(\lambda_\ell)
\left(\prod_{m=1\atop{m\ne \ell}}^{N+1}g(\lambda_\ell, \lambda_m)\right)
\left(\prod_{m=1\atop{m\ne \ell}}^{N+1}h(\lambda_\ell, \lambda_m)\right)
\non
&&\times\bra \prod_{j=1}^NC(\mu_j)
\prod_{m=1\atop{m\ne \ell}}^{N+1}B(\lambda_m)
\ket.
\label{form1}
\ea
Let us notice that the form factor is symmetric function of all the $\lambda_j$ because
$$[B(\lambda_j),B(\lambda_k)]=0.$$

 We now need to calculate the scalar product between the eigenvector and
 non-eigenvector
\be
\bra \prod_{j=1}^NC(\mu_j)
\prod_{m=1\atop{m\ne \ell}}^{N+1}B(\lambda_m)\ket,
\ee
where $\mu_j$ satisfy the Bethe equations, but $\lambda_m$ do not. It can be
done by the following theorem.
\begin{thm}
The following determinant representation holds for
such scalar products:
\begin{eqnarray}
&&\bra\prod_{j=1}^NC(\mu_j) \prod_{j=1}^N B(\lambda_j)
\vert 0 \rangle=
\left\{\prod_{j=1}^Nd(\mu_j)d(\lambda_j)\right\}\nonumber \\
&\times&
\left\{\prod_{N\geq j >k \geq 1}g(\lambda_j,\lambda_k)
g(\mu_k,\mu_j)\right\}
\left\{\prod_{j,k=1}^Nh(\mu_j,\lambda_k)\right\}
\det\left(M_{jk}\right),
\end{eqnarray}
where
\begin{eqnarray}
M_{jk}=\frac{g(\mu_k,\lambda_j)}{h(\mu_k,\lambda_j)}
-\frac{a(\lambda_j)}{d(\lambda_j)}
\frac{g(\lambda_j,\mu_k)}{h(\lambda_j,\mu_k)}
\prod_{m=1}^N\frac{f(\lambda_j,\mu_m)}{f(\mu_m,\lambda_j)}.
\end{eqnarray}
Here the spectral parameters $\{\mu_j\}$ satisfy the Bethe Ansatz equations 
(\ref{BETHEeq}).
The spectral parameters $\{\lambda_j\}$ are free and do not satisfy any
equations.
\end{thm}

This theorem was proved in  \cite{S}.

For the scalar product, which appears in the expression for the form factor 
we get
\be
\begin{array}{l}{\dis
\bra \prod_{j=1}^NC(\mu_j)
\prod_{m=1\atop{m\ne \ell}}^{N+1}B(\lambda_m)\ket}\\
=
\begin{array}[t]{l}
{\dis
\prod_{N\geq j>k\geq 1}
g(\mu_j,\mu_k)
\cdot\prod_{N+1\geq j>k\geq 1\atop{j\neq\ell,
\ k\neq \ell}}g(\lambda_k, \lambda_j)
\cdot
\prod_{j=1}^N \prod_{m=1\atop{m\ne \ell}}^{N+1}
h(\mu_j,\lambda_m)}\\
\\
{\dis
\times\prod_{j=1}^N d(\mu_j)
\cdot\prod_{m=1\atop{m\ne \ell}}^{N+1}d(\lambda_m)
\cdot {\det}_{N}M^{(\ell)}}.
\end{array}\end{array}
\label{SS1}
\ee
Here the entries of the $N\times N$ matrix $M^{(\ell)}$ are
\be
M^{(\ell)}_{jk}=t(\mu_k,\lambda_j)
-r(\lambda_j)t(\lambda_j,\mu_k)
\cdot \prod_{m=1}^N \frac{f(\lambda_j,\mu_m)}{
f(\mu_m,\lambda_j)},
\qquad
\begin{array}{c}
j=1,\dots,\ell-1,\ell+1,\dots,N+1\\
k=1,\dots,N
\end{array}
\label{Mell1}
\ee
Let us recall that
$$
t(\lambda, \mu)=\frac{(ic)^2}{(\lambda-\mu)(\lambda-\mu+ic)}
\qquad\mbox{and}\qquad r(\lambda)=\frac{a(\lambda)}{d(\lambda)}.
$$
Remember that the Bethe equations give:
\be
r(\lambda_j)=\prod_{p=1}^{N+1} \frac{h(\lambda_p,\lambda_j)}{
h(\lambda_j,\lambda_p)}
\quad
\mbox{and}
\quad
\prod_{m=1}^{N} \frac{f(\lambda_j,\mu_m)}{
f(\mu_m,\lambda_j)}
= (-1)^{N}
\prod_{m=1}^{N} \frac{h(\lambda_j,\mu_m)}{
h(\mu_m,\lambda_j)},
\ee
or equivalently,
\be
a(\lambda_\ell)\prod_{m=1}^{N+1}h(\lambda_\ell, \lambda_m)
=
d(\lambda_\ell)\prod_{m=1}^{N+1} h(\lambda_m, \lambda_\ell).
\ee
Expression \Eq{Mell1} becomes
\be
M^{(\ell)}_{jk}=t(\mu_k,\lambda_j)
-t(\lambda_j,\mu_k)
\left(
\prod_{p=1}^{N+1}\frac{h(\lambda_p,\lambda_j)}{
h(\lambda_j,\lambda_p)}
\right)\cdot
\left(
\prod_{m=1}^{N}\frac{h(\lambda_j,\mu_m)}{
h(\mu_m,\lambda_j)}
\right).
\ee
Using the obvious equality
\ba
\prod_{m=1\atop{m\ne \ell}}^{N+1}g(\lambda_\ell, \lambda_m)
&=&
\left(\prod_{m=1}^{\ell-1}g(\lambda_\ell, \lambda_m)\right)
\left(\prod_{m=\ell+1}^{N+1}g(\lambda_\ell, \lambda_m)\right)
\non
&=&
(-1)^{\ell-1}
\left(\prod_{m=1}^{\ell-1}g(\lambda_m, \lambda_\ell)\right)
\left(\prod_{m=\ell+1}^{N+1}g(\lambda_\ell, \lambda_m)\right),
\ea
 and substituting \Eq{SS1} into \Eq{form1}, we have
\ba {\dis
F_N=-i\sqrt{c}\sum_{\ell=1}^{N+1}(-1)^{\ell-1}}&&
{\dis
	\prod_{N\geq j>k\geq 1}g(\mu_j, \mu_k)
	\prod_{N+1\geq j>k\geq 1}g(\lambda_k, \lambda_j)
}
        \non
&&      \non
&&{\dis
        \times\prod_{m=1}^{N+1} h(\lambda_m,\lambda_\ell)
	\prod_{j=1}^N\prod_{m=1\atop{m\ne \ell}}^{N+1}h(\mu_j,\lambda_m)
}       \non
&&{\dis
        \times\prod_{j=1}^{N}d(\mu_j)
	\prod_{m=1}^{N+1}d(\lambda_m)
}       \cdot{\det}_N M^{(\ell)}.
\label{form2}
\ea
One can rewrite the determinant ${\det}_N M^{(\ell)}$ as
\be
{{\det}}_NM^{(\ell)}=
\left(
	\prod_{m=1}^N \prod_{j=1\atop{j\ne\ell}}^{N+1}
	\frac{1}{h(\mu_m,\lambda_j)}
\right)
\cdot
\left(
	\prod_{p=1}^{N+1}\prod_{j=1\atop{j\ne\ell}}^{N+1}
	h(\lambda_p,\lambda_j)
\right)
\cdot
{\det}_N S^{(\ell)},
\label{M2S}
\ee
where
\be
S_{jk}^{(\ell)}=
t(\mu_k,\lambda_j)
\frac{\prod\limits_{m=1}^N h(\mu_m,\lambda_j)}{
\prod\limits_{p=1}^{N+1}h(\lambda_p,\lambda_j)}
-
t(\lambda_j,\mu_k)
\frac{\prod\limits_{m=1}^N h(\lambda_j,\mu_m)}{
\prod\limits_{p=1}^{N+1}h(\lambda_j,\lambda_p)},
\qquad
\begin{array}{c}
j=1,\dots,\ell-1,\ell+1,\dots,N+1\\
k=1,\dots,N
\end{array}
\label{S1}
\ee
Let us substitute \Eq{M2S} into \Eq{form2},
\ba
F_N&=&
{\dis
	-i\sqrt{c}
	\prod_{N\geq j>k\geq 1}g(\mu_j, \mu_k)
	\prod_{N+1\geq j>k\geq 1}g(\lambda_k, \lambda_j)
	\prod_{m=1}^{N+1}\prod_{j=1}^{N+1}h(\lambda_m,\lambda_j)
}\non
&&\non
&&{\dis
        \times
	\left(
	\sum_{\ell=1}^{N+1}(-1)^{\ell+1}
        {\det}_N S^{(\ell)}
	\right)
	\cdot
	\prod_{j=1}^{N}d(\mu_j)
        \prod_{m=1}^{N+1}d(\lambda_m).
}
\label{form3}
\ea
In order to  simplify this expression let us study
\be
\cM=\cMi\lambda\equiv\sum_{\ell=1}^{N+1}(-1)^{\ell-1}
{\det}_N S^{(\ell)}.
\label{Mi1}
\ee
Notice that $\cM$ is an antisymmetric function of all
$\{\lambda_j\}$ because $F_N$ is symmetric
and the product of functions
$g(\lambda_k,\lambda_j)$
is antisymmetric.  In particular,
\be
\cMi{\lambda}=0\quad
\mbox{if}
\quad
\lambda_j=\lambda_k.
\ee
${\det} S^{(\ell)}$ can be obtained from
${\det} S^{(N+1)}$ by replacing $\lambda_{\ell}$
and $\lambda_{N+1}$.
This is a special case of a permutation
\be
(\lambda_1,\cdots,\lambda_\ell,\cdots,
\lambda_N,\lambda_{N+1})
\longrightarrow
(\lambda_1,\cdots,\lambda_{N+1},\cdots,
\lambda_N,\lambda_\ell).
\ee
Since $(-1)^{\ell-1}$ is the parity of this permutation,
\ba
\cM\{\lambda\} &=& \sum_{{\scriptstyle \mbox{\scriptsize
Permutation of all }\{\lambda_{N+1}\}}}
(-1)^{P}\prod_{j=1}^N S_{P(j)j}\non
&=& \left(1+\frac{\partial}{\partial \alpha}\right)
\left.{\det}_{N} (S_{jk}-\alpha S_{N+1,k})\right|_{\alpha=0}.
\label{Mi2}
\ea
Here $S_{jk}$ means $S_{jk}^{(N+1)}$ from \Eq{S1} and
${\det} S_{jk}$ is the term $\ell=N+1$ in \Eq{Mi1},
$$
-\frac{\partial}{\partial \alpha}
\left.{\det}_{N} (S_{jk}-\alpha S_{N+1,k})\right|_{\alpha=0}
$$
is the sum of $N$ terms where each of them  differs from ${\det} S_{jk}$
by the replacement of the $\ell$-th line
(corresponding to $\lambda_\ell$) by the $(N+1)$-th line.
We can use the  expression \Eq{Mi2} to simplify  the form
factor \Eq{form3}
\ba
F_N&=&
{\dis
	-i\sqrt{c}
	\prod_{N\geq j>k\geq 1}g(\mu_j, \mu_k)
	\prod_{N+1\geq j>k\geq 1}g(\lambda_k, \lambda_j)
	\prod_{m=1}^{N+1}\prod_{j=1}^{N+1}h(\lambda_m,\lambda_j)
}\non
&&\non
&&{\dis
        \times
	\prod_{j=1}^{N}d(\mu_j)
	\prod_{m=1}^{N+1}d(\lambda_m)
	\cdot\cMi{\lambda}.
}
\label{form4}
\ea

The complex conjugate of form factor is
\be
\overline{F}_N=\bra \prod_{j=1}^{N+1} C(\lambda_j)\psi^\dagger(0,0)
\prod_{j=1}^NB(\mu_j)\ket.
\ee
 Remember that $c$ and all $\lambda$, $\mu$ are real.
Therefore complex conjugation gives
\be
\begin{array}{l}
\overline{g(\lambda,\mu)}=g(\mu,\lambda),\quad
\overline{f(\lambda,\mu)}=f(\mu,\lambda),\quad
\overline{h(\lambda,\mu)}=h(\mu,\lambda),\\
\overline{t(\lambda,\mu)}=t(\mu,\lambda),\quad
\overline{a(\lambda)} = d(\lambda)=a^{-1}(\lambda).
\end{array}
\ee
So we have
$$
\overline{S}_{jk}=-S_{jk},
$$
and for even $N$ 
\be
\overline{\cMi{\lambda}}=(-1)^N \cMi{\lambda}
=\cMi{\lambda}.
\ee

Hence for the complex conjugated form factor $\overline F_N$ we get
\ba
\overline F_N&=&
{\dis
	i\sqrt{c}
	\prod_{N\geq j>k\geq 1}g(\mu_k, \mu_j)
	\prod_{N+1\geq j>k\geq 1}g(\lambda_j, \lambda_k)
	\prod_{m=1}^{N+1}\prod_{j=1}^{N+1}h(\lambda_j,\lambda_m)
}\non
\non
&&{\dis
        \times
	\prod_{j=1}^{N}a(\mu_j)
	\prod_{m=1}^{N+1}a(\lambda_m)
	{\cMi{\lambda}}.
}
\label{form5}
\ea

For the correlation function the quantity $|F_N|^2$ is important:
\ba
F_N\overline{F}_N & = &
c\left(\prod_{j=1}^N a(\mu_j)d(\mu_j)\right)
\left(\prod_{m=1}^{N+1} a(\lambda_m)d(\lambda_m)\right)
\cdot
\left(\prod_{j=1,k=1\atop{j\neq k}}^N g(\mu_j,\mu_k)\right)
\non
&&\times
\left(\prod_{j=1,k=1\atop{j\neq k}}^{N+1} g(\lambda_j,\lambda_k)\right)
\cdot
\left(\prod_{j=1}^{N+1}\prod_{k=1}^{N+1} h(\lambda_j,\lambda_k)\right)^2
(\cMi{\lambda})^2.
\ea
or,
\be
\begin{array}{l}
{\dis
	\frac{F_N\overline{F}_N}{
        \bra\prod\limits_{j=1}^NC(\mu_j)
        \prod\limits_{j=1}^N B(\mu_j)\ket\cdot
        \bra\prod\limits_{j=1}^{N+1}C(\lambda_j)
        \prod\limits_{j=1}^{N+1}
	B(\lambda_j)\ket}
}\\
\\
{\dis
	=c^{-2N}\frac{
        \left(\prod\limits_{j=1}^{N+1}\prod\limits_{k=1}^{N+1}
	h(\lambda_j,\lambda_k)\right)
	\cdot
	(\cMi{\lambda})^2
	}{
        \left(\prod\limits_{j=1}^N
        \prod\limits_{k=1}^N h(\mu_j,\mu_k)\right)
        {\det}_N\frac{\partial \varphi_j}{\partial \mu_k}
        {\det}_{N+1}\frac{\partial \tilde\varphi_j}{\partial \lambda_k}}
}
\end{array}
\ee
This formula gives us $|F_N|^2$ at $x=t=0$. Also it is easy to
``switch'' on space and time dependence using the formula
\Eq{cc1}. Note that
$a(\lambda)d(\lambda)=1$ for nonlinear Schr\"odinger equation.
Therefore the correlation function becomes
\be\label{corr1}
\langle \psi(0,0)\psi^\dagger(x,t)\rangle_N
=\left\{\frac{e^{-iht}c^{-2N}}{(\prod\limits_{j=1}^N
\prod\limits_{k=1}^N h(\mu_j,\mu_k))\cdot
{\det}_N\frac{\partial \varphi_j}{\partial \mu_k}}\right\}
\sum_{\{\lambda\}} \frac{\Delta(\{\lambda\})}
{{\det}_{N+1}\frac{\partial \tilde\varphi_j}{\partial \lambda_k}}.
\ee
Here we used the new notation
\be\label{numerator}
\Delta(\{\lambda\})=
\left(\prod_{k=1}^{N+1}\prod_{j=1}^{N+1}h(\lambda_j,\lambda_k)\right)
(\cMi{\lambda})^2
e^{\left\{
\sum_{j=1}^{N+1}\tau(\lambda_j)-\sum_{m=1}^N\tau(\mu_m)\right\}},
\ee
where
\be\label{tau}
\tau(\lambda)=it\lambda^2-ix\lambda.
\ee

\section{The idea of summation with respect to $\lambda$}

Now let us consider the sum with respect to all 
$\left\{\lambda\right\}_{N+1}$ in \Eq{corr1}
$$
\sum_{\left\{\lambda\right\}_{N+1}}
\frac{\Delta(\{\lambda\})}{{\det}_{N+1}\left(
\frac{\partial \tilde\varphi_j}{\partial \lambda_k}
\right)}
$$
The idea of summation is the same as that which we used for impenetrable
bosons (free fermionic case) \cite{K.S.1} 
(see also Section XIII.5 of \cite{K.B.I.}). The factor
$(\cMi{\lambda})^2$ entering  the r.h.s. of \Eq{numerator} contains
$(N+1)!$ terms, all of them give the same contribution to the sum. So
we can replace one of determinants $\cMi{\lambda}$ by the product
$\prod_{j=1}^NS_{jj}$
\ba
&&\sum_{\left\{\lambda\right\}_{N+1}}
\frac{\Delta(\{\lambda\})}{{\det}_{N+1}\left(
\frac{\partial \tilde\varphi_j}{\partial \lambda_k}
\right)}\non
&&=(N+1)!
\sum_{\left\{\lambda\right\}_{N+1}}
\prod_{m=1}^N e^{-\tau(\mu_m)}
\cdot\left(
{\det}_{N+1}\frac{\partial \tilde\varphi_j}{\partial \lambda_k}
\right)^{-1}\non
&&\times
\left(\prod_{k=1}^{N+1}\prod_{j=1}^{N+1}h(\lambda_j,\lambda_k)\right)
\prod_{j=1}^{N+1}e^{\tau(\lambda_j)}
\left(\prod_{j=1}^N S_{jj}\right)\cM\{\lambda\}.
\label{corr3}
\ea
The sum with respect to all $\left\{\lambda\right\}_{N+1}$
means the sum with respect to all ordered sets of integers
$\left\{n_j\right\}$ from \Eq{tildevarphi}.
We also can admit $n_j=n_k$ because it leads to 
$\lambda_j=\lambda_k$ which does not contribute to 
\Eq{corr3} because of the antisymmetry of $\cM\{\lambda\}$.
The factor $(N+1)!$ is absorbed as
\be
(N+1)!\sum_{\left\{n_j\right\}}
=
\prod_{i=1}^{N+1}\sum_{n_i=-\infty}^\infty.
\ee
The correlation function becomes
\ba{\dis
\langle \psi(0,0)\psi^\dagger(x,t)\rangle_N}
&=&{\dis \frac{e^{-iht}c^{-2N}
\prod_{m=1}^N e^{-\tau(\mu_m)}
}{(\prod_{j=1}^N
\prod_{k=1}^N h(\mu_j,\mu_k))\cdot
{\det}_N\frac{\partial \varphi_j}{\partial \mu_k}}}
\non
&&{\dis\times
\sum_{n_1\cdots n_{N+1}}\left(
\frac{\widetilde\Delta(\{\lambda\})}{
{\det}_{N+1}\frac{\partial \tilde\varphi_j}{\partial \lambda_k}
}\right).}
\label{corr4}
\ea
with
\ba{\dis
\widetilde\Delta(\{\lambda\})}&=&
{\dis \left(
\prod_{j=1}^{N+1}\prod_{k=1}^{N+1}h(\lambda_j,\lambda_k)\right)
\prod_{j=1}^{N+1}e^{\tau(\lambda_j)}}\non
& \times &{\dis
\left.\left(1+\frac{\partial}{\partial \alpha}\right)
{\det}_N(S_{jj}S_{jk}-\alpha S_{jj}S_{N+1\,k})\right|_{\alpha=0}}.
\label{num1}
\ea

The main difference between the free fermionic case (coupling constant
$c\to+\infty$) and the non-free fermionic case is that in the former
it is possible to solve the Bethe equations \Eq{BETheeq} explicitly. On
the contrary this is not possible for penetrable bosons.

Our approach is based on the formula
$$
 {\dis
\sum_{n_1\cdots n_{N+1}}\left( \frac{\widetilde\Delta(\{\lambda\})}{ 
{\det}_{N+1}\frac{\partial \tilde\varphi_j}{\partial \lambda_k}
}\right)
=\sum_{n_1\cdots n_{N+1}}
\int\limits_{-\infty}^\infty d^{N+1}\lambda\, 
\widetilde\Delta(\{\lambda\})
\prod_{j=1}^{N+1}\delta(\tilde\varphi_j(\lambda)-2\pi n_j)}.
$$
Remember that $\det\partial \tilde\varphi_j/\partial \lambda_k>0$. We shall
also use the Poisson formula
\be
\sum_{n=-\infty}^\infty \delta(x-2\pi n)
=
\frac{1}{2\pi}\sum_{k=-\infty}^\infty
e^{ikx}.
\label{summation}
\ee
So we have
\ba && {\dis
\sum_{n_1\cdots n_{N+1}}\left( \frac{\widetilde\Delta(\{\lambda\})}{ 
{\det}_{N+1}\frac{\partial \tilde\varphi_j}{\partial \lambda_k}
}\right)
=\sum_{n_1\cdots n_{N+1}}
\int\limits_{-\infty}^\infty d^{N+1}\lambda\, 
\widetilde\Delta(\{\lambda\})
\prod_{j=1}^{N+1}\delta(\tilde\varphi_j(\lambda)-2\pi n_j)}\non
&&{\dis \hskip1cm
=\sum_{n_1\cdots n_{N+1}}
\left(\frac{1}{2\pi}\right)^{N+1}
\int\limits_{-\infty}^\infty d^{N+1}\lambda\,
\widetilde\Delta(\{\lambda\})
\prod_{j=1}^{N+1}e^{in_j\tilde\varphi_j(\lambda)}}\non
&&{\dis \hskip1cm
=\sum_{n_1\cdots n_{N+1}}
\left(\frac{1}{2\pi}\right)^{N+1}
\int\limits_{-\infty}^\infty d^{N+1}\lambda
\widetilde\Delta(\{\lambda\})
\,\,\prod_{j=1}^{N+1}e^{iL\lambda_jn_j}
\left(\prod_{k=1}^{N+1} \frac{h(\lambda_k,\lambda_j)}{
h(\lambda_j,\lambda_k)}\right)^{n_j}.}
\label{int1}
\ea
Thus we get  the following representation for the correlation function
\ba
&&{\dis\langle \psi(0,0)\psi^\dagger(x,t)\rangle_N}
={\dis \frac{e^{-iht}c^{-2N}
\prod_{m=1}^N e^{-\tau(\mu_m)}
}{(\prod_{j=1}^N
\prod_{k=1}^N h(\mu_j,\mu_k))\cdot
{\det}_N\frac{\partial \varphi_j}{\partial \mu_k}}}
\non
&&{\dis \hskip1cm
\times\sum_{n_1\cdots n_{N+1}}
\left(\frac{1}{2\pi}\right)^{N+1}
\int\limits_{-\infty}^\infty d^{N+1}\lambda
\widetilde\Delta(\{\lambda\})
\,\,\prod_{j=1}^{N+1}e^{iL\lambda_jn_j}
\left(\prod_{k=1}^{N+1} \frac{h(\lambda_k,\lambda_j)}{
h(\lambda_j,\lambda_k)}\right)^{n_j}.}
\label{correlator}
\ea
\vskip1cm
\section{Quantum Dual Fields}

In this section we introduce the auxiliary Fock space and auxiliary
Bose  fields
$\phi_0(\lambda)$, $\phi_1(\lambda)$, $\phi_2(\lambda)$
$\phi_{A_j}(\lambda)$ and $\phi_{D_j}(\lambda)$ ($j=1,2$).
Further we shall call these operators  dual fields \cite{KOR1}
(see also Section IX.5 of \cite{K.B.I.}).
Dual fields help us to rewrite double products in terms of single
products.

By definition any operator $\phi_a(\lambda)$
($a=0$, $1$, $2$, $A_1$, $A_2$, $D_1$, $D_2$)  is  the sum 
of two operators: "momentum" $p(\lambda)$ and "coordinate" 
$q(\lambda)$.  

\be\label{defdualfields}
 \begin{array}{lcl}{\dis 
\phi_0(\lambda)=q_0(\lambda)+p_0(\lambda);}&&\\
{\dis
\phi_{A_j}(\lambda)=q_{A_j}(\lambda)+p_{D_j}(\lambda);}
&\qquad&{\dis
\phi_{D_j}(\lambda)=q_{D_j}(\lambda)+p_{A_j}(\lambda);}\\
{\dis
\phi_1(\lambda)=q_1(\lambda)+p_2(\lambda);}
&\qquad &{\dis
\phi_2(\lambda)=q_2(\lambda)+p_1(\lambda).}
\end{array}
\ee
All operators "momenta"  $p(\lambda)$ annihilate  the vacuum vector
$\ketd$, all operators $q(\lambda)$ annihilate the dual vacuum $\brad$:
$$
p_a(\lambda)\ketd=0, \quad \brad q_a(\lambda)=0, \qquad
\mbox{for all $a$},\qquad  (0|0)=1.
$$
The only nonzero commutation relations are
\be\label{commutators}\left\{
\begin{array}{lcl}
{\dis
[p_0(\lambda),q_0(\mu)]=\ln(h(\lambda,\mu)h(\mu,\lambda));}
&&\\
{\dis
[p_{D_j}(\lambda),q_{D_k}(\mu)]=\delta_{jk}\ln h(\lambda,\mu);}
&\qquad&
{\dis
[p_{A_j}(\lambda),q_{A_k}(\mu)]=\delta_{jk}\ln 
h(\mu,\lambda);}\\
{\dis
[p_1(\lambda),q_1(\mu)]
 = \ln\frac{h(\lambda,\mu)}{h(\mu,\lambda)};}
&\qquad& {\dis
[p_2(\lambda),q_2(\mu)]
=\ln\frac{h(\mu,\lambda)}{h(\lambda,\mu)}.}
\end{array} \right.
\ee
(We remind the reader that $h(\lambda,\mu)=(\lambda-\mu+ic)/ic$).
 The realization of these operators
as linear combinations of canonical Bose fields is given in
Appendix B.

It is easy to check that all dual fields commute with each other
$$
\qquad [\phi_a(\lambda),\phi_b(\mu)]=0,
$$
where $a,b$ run through the all possible indices. Using this property
we can define  functions of operators
${\cal F}(\{e^{\phi_a(\lambda)}\})$. One should understand such 
expression, for example as a power series over $\{e^{\phi_a(\lambda)}\}$. The 
following simple formul\ae\ are useful:

\ba
&&\label{mean1}{\dis e^{p_a(\lambda)}e^{q_a(\mu)}=
e^{q_a(\mu)}e^{p_a(\lambda)}e^{[p_a(\lambda),q_a(\mu)]}},\\ 
&&\label{mean2} {\dis  
\brad\prod_{j=1}^{M_1}e^{\alpha_j p_a(\lambda_j)}
\prod_{k=1}^{M_2}e^{\beta_k q_a(\mu_k)}\ketd=
\prod_{j=1}^{M_1}\prod_{k=1}^{M_2}e^{\alpha_j\beta_k
[p_a(\lambda_j),q_a(\mu_k)]},}\\
&&\label{mean3}{\dis
\prod_{j=1}^{M}e^{\beta_jp_a(\lambda_j)}
{\cal F}\left(e^{\phi_a(\mu)}\right)\ketd=
{\cal F}\left(e^{\phi_a(\mu)}\prod_{j=1}^{M}e^{\beta_j
[p_a(\lambda_j),q_a(\mu)]}\right)\ketd,}\\
&&\label{mean4}{\dis
{\cal F}\left(e^{p_a(\mu)}\right)
\prod_{j=1}^{M}e^{\beta_j\phi_a(\lambda_j)}\ketd=
\prod_{j=1}^{M}e^{\beta_j\phi_a(\lambda_j)}\ketd
{\cal F}\left(\prod_{j=1}^{M}e^{\beta_j
[p_a(\mu),q_a(\lambda_j)]}\right).}
\ea
Here $\{\lambda\},\{\mu\},\{\beta\},\{\alpha\}$ are arbitrary complex 
numbers, ${\cal F}$ is a function. One can easily prove
these formul\ae.

Let us define the very important dual field $\psi(\lambda)$ as
\be\label{psi}
\psi(\lambda)=\phi_{0}(\lambda)+\phi_{A_1}(\lambda)
+\phi_{D_2}(\lambda)+\phi_{2}(\lambda).
\ee

\begin{thm}~~The correlation function \Eq{correlator} can be presented as the following vacuum expectation value in auxiliary Fock space
\ba{\dis
\langle \psi(0,0)\psi^\dagger(x,t)\rangle_N}
&=&{\dis \frac{e^{-iht}c^{-2N}}
{{\det}_N\frac{\partial \varphi_j}{\partial \mu_k}}
\brad\prod_{m=1}^{N}\left(e^{p_0(\mu_m)}
e^{p_1(\mu_m)}\right) }\non
&\times &
{\dis 
\frac{1}{(2\pi)^{N+1}}\sum_{n_1\cdots n_{N+1}}
\int\limits_{-\infty}^{\infty}\,d^{N+1}\lambda
\left(\hat \gamma_1(\lambda_{N+1})
+\frac{\partial}{\partial\alpha}\right)}
\non
&\times &{\dis
{\det}_N\left(\hS_{jj}\hS_{jk}\hat\gamma_1(\lambda_j)
\hat\gamma_2(\mu_j)\hat\gamma_2(\mu_k)\right.}\non
&-&{\dis\left.\left.
\alpha \hS_{jj} \hS_{N+1\,k}
\hat\gamma_1(\lambda_j) \hat\gamma_1(\lambda_{N+1})
\hat\gamma_2(\mu_j)\hat\gamma_2(\mu_k)\right)
\right|_{\alpha=0}\ketd.}\label{theorema}
\ea

where
\be
\label{hS}
\hS_{jk}=t(\mu_k,\lambda_j)e^{-\phi_{D_1}(\lambda_j)}-
t(\lambda_j,\mu_k)e^{-\phi_{A_2}(\lambda_j)},
\ee
and
$$
\begin{array}{rcl}
\hat\gamma_1(\lambda_j)&=&e^{iL\lambda_jn_j+
\tau(\lambda_j)+\psi(\lambda_j)
+n_{j}\phi_1(\lambda_{j})},\\
\hat\gamma_2(\mu)&=&e^{-\frac{1}{2}(\tau(\mu)+\psi(\mu))}.
\end{array}
$$
\end{thm}

{\it Proof.}
Let us move factors $\hat\gamma_1(\lambda)$ and
$\hat\gamma_2(\mu)$ out of the determinant in \Eq{theorema}. 
In the r.h.s. of
\Eq{theorema} we get

\ba\label{proof1}
\quad &\quad &{\dis 
\left(1+\frac{\partial}{\partial\alpha}\right)
\brad\prod_{m=1}^{N}\left[e^{p_0(\mu_m)}
e^{p_1(\mu_m)}\right] }\non
&\times &{\dis
\prod_{m=1}^{N+1}\left[e^{iL\lambda_mn_m+
\tau(\lambda_m)+\psi(\lambda_m)
+n_{m}\phi_1(\lambda_{m})}\right]
\prod_{m=1}^{N}\left[e^{-\tau(\mu_m)-\psi(\mu_m)}\right]
}\non                
&\times &{\dis \left.
{\det}_N\left(\hS_{jj}\hS_{jk}-
\alpha \hS_{jj} \hS_{N+1\,k}\right)
\right|_{\alpha=0}\ketd.}
\ea

Using \Eq{mean2}  we find
\ba\label{proof2}&&{\dis
\brad\prod_{m=1}^{N}\left[e^{p_0(\mu_m)}
e^{p_1(\mu_m)}\right] 
\prod_{m=1}^{N+1}\left[e^{\phi_0(\lambda_m)
+n_{m}\phi_1(\lambda_{m})+\phi_2(\lambda_m)}\right]
\prod_{m=1}^{N}\left[e^{-\phi_0(\mu_m)-\phi_2(\mu_m)}\right]
\ketd}\non
&&{\dis \mbox{\hskip4cm}
=\frac{\prod_{j=1}^{N+1}\prod_{k=1}^{N+1}h(\lambda_j,\lambda_k)}
{\prod_{j=1}^{N}\prod_{k=1}^{N}h(\mu_j,\mu_k)}
\prod_{j=1}^{N+1}\prod_{k=1}^{N+1}
\left(\frac{h(\lambda_k,\lambda_j) }{h(\lambda_j,\lambda_k)}
\right)^{n_j}.}
\ea

Using \Eq{mean3} we obtain
\ba\label{proof3}&&{\dis
\brad\prod_{m=1}^{N+1}\left[e^{\phi_{A_1}(\lambda_m)
+\phi_{D_2}(\lambda_{m})}\right]
\prod_{m=1}^{N}\left[e^{-\phi_{A_1}(\mu_m)-\phi_{D_2}(\mu_m)}\right]
{\det}_N\left(\hS_{jj}\hS_{jk}-\alpha\hS_{jj}\hS_{N+1\,k}\right)
\ketd}
\non
&&{\dis \mbox{\hskip5cm}
={\det}_N\left(S_{jj}S_{jk}-\alpha S_{jj}S_{N+1\,k}\right).}
\ea
Combining \Eq{proof2} and \Eq{proof3} we get the r.h.s. of
\Eq{correlator} with $\widetilde\Delta(\{\lambda\})$ defined in
\Eq{num1}. The theorem is proved.

Now we can rewrite the r.h.s. for \Eq{theorema} as follows
\ba\label{summation1}
&&{\dis 
\frac{1}{(2\pi)^{N+1}}\sum_{n_1\cdots n_{N+1}}
\int\limits_{-\infty}^{\infty}\,d^{N+1}\lambda
\left(\hat \gamma_1(\lambda_{N+1})
+\frac{\partial}{\partial\alpha}\right)}\non
&&\times{\dis 
\left.{\det}_N\left(\hS_{jj}\hS_{jk}\hat\gamma_1(\lambda_j)
\hat\gamma_2(\mu_j)\hat\gamma_2(\mu_k)
-\alpha \hS_{jj} \hS_{N+1\,k}
\hat\gamma_1(\lambda_j) \hat\gamma_1(\lambda_{N+1})
\hat\gamma_2(\mu_j)\hat\gamma_2(\mu_k)\right)
\right|_{\alpha=0}}\non
&&={\dis  
\left(\frac{1}{2\pi}\sum_{n_{N+1}}
\int\limits_{-\infty}^{\infty}\,d\lambda_{N+1}
\hat \gamma_1(\lambda_{N+1})
+\frac{\partial}{\partial\alpha}\right)}\non
&&\times{\dis 
{\det}_N\left(
\frac{1}{2\pi}\sum_{n_j}
\int\limits_{-\infty}^{\infty}\,d\lambda_j
\hS_{jj}\hS_{jk}\hat\gamma_1(\lambda_j)
\hat\gamma_2(\mu_j)\hat\gamma_2(\mu_k)\right.}\non
&&-{\dis \left.\left.
\alpha
\frac{1}{4\pi^2}\sum_{n_j,n_{N+1}}
\int\limits_{-\infty}^{\infty}\,d\lambda_{N+1}d\lambda_j
 \hS_{jj} \hS_{N+1\,k}
\hat\gamma_1(\lambda_j) \hat\gamma_1(\lambda_{N+1})
\hat\gamma_2(\mu_j)\hat\gamma_2(\mu_k)\right)
\right|_{\alpha=0}.}
\ea

In this formula  we can perform the summation over integer 
$\{n_j\}$. Indeed, $n_j$ enters only in function 
$\hat\gamma_1(\lambda_j)$:  
$$ 
\hat\gamma_1(\lambda_j)=e^{iL\lambda_jn_j+
\tau(\lambda_j)+\psi(\lambda_j)
+n_{j}\phi_1(\lambda_{j})}.
$$

Recall that all dual fields commute with each other:
$$
[\phi_a(\lambda),\phi_b(\mu)]=0.
$$
This means that we can treat operators $\phi_a(\lambda)$ as 
diagonal operators. Due to the formula \Eq{hermit} from Appendix B
we can consider operator $i\phi_1(\lambda)$ as real function of $\lambda$.
 Hence  we can use formula \Eq{summation} to sum up with respect to
$n_j$
\be\label{Poisson}
\frac{1}{2\pi}\sum_{n=-\infty}^{\infty}
e^{in(L\lambda-i\phi_1(\lambda))}=
\sum_{n=-\infty}^{\infty}\delta(L\lambda-i\phi_1(\lambda)-2\pi n).
\ee
It means that  $\lambda=\lambda_n$, where $\lambda_n$ is a root of the equation
\be\label{mainequation}
L\lambda_n-2\pi n =i\phi_1(\lambda_n).
\ee

The expression \Eq{mainequation} is an operator equality,
which is defined only on vectors of the form 
$\prod\limits_{m}e^{\phi_2(\lambda_m)}|0)$. Therefore
 one should understand this
equation in the sense of mean value:
\be\label{understand}
(0|(L\lambda_n-i\phi_1(\lambda_n)-2\pi n)\prod_{m}
e^{\phi_2(\lambda_m)}|0)=0,
\ee
where $\{\lambda_m\}$ are arbitrary real parameters. Then we can rewrite
equation \Eq{understand}:                                                         
\be\label{understand1}
L\lambda_n-2\pi n=i\sum_{m}
\ln\frac{h(\lambda_m,\lambda_n)}{h(\lambda_n,\lambda_m)}=
i\sum_{m}
\ln\left(\frac{\lambda_m-\lambda_n+ic}{\lambda_n-\lambda_m+ic}\right).
\ee
The r.h.s. of \Eq{understand1} is a real bounded function 
of $\lambda_n$. Moreover it is a decreasing function of $\lambda_n$, because
$$
\frac{\partial}{\partial\lambda_n}i\sum_{m}
\ln\left(\frac{\lambda_m-\lambda_n+ic}{\lambda_n-\lambda_m+ic}\right)=
-\sum_m\frac{2c}{(\lambda_n-\lambda_m)^2+c^2}<0.
$$
The l.h.s. of \Eq{understand1} is a linear increasing
function of $\lambda_n$, hence
equation \Eq{understand1} has one real solution and this solution is unique.
Also we have
\be\label{positive}
(0|(L-i\phi'_1(\lambda_n))\prod_{m}
e^{\phi_2(\lambda_m)}|0)=L+
\sum_m\frac{2c}{(\lambda_n-\lambda_m)^2+c^2}>0.
\ee
Therefore, one can write
\be\label{Poisson1}
\delta(L\lambda-i\phi_1(\lambda)-2\pi n)=
\frac{\delta(\lambda-\lambda_n)}{L-i\phi'_1(\lambda)},
\ee
where $\lambda_n$ is solution of equation \Eq{mainequation}.

Later we shall use notation
\be\label{hrho}
\hrp(\lambda)=1-\frac iL\phi'_1(\lambda).
\ee
We now arrive at the following formula for the correlation function
in finite volume

\ba
\langle \psi(0,0)\psi^\dagger(x,t)\rangle_N
& = &
\frac{e^{-iht}c^{-2N}}{
{\det}_N\left(\frac{\partial \varphi_j}{\partial \mu_k}\right)}
\brad\prod\limits_{m=1}^N\left(e^{p_0(\mu_m)}e^{p_1(\mu_m)}\right)
\non
&&\times\left.
 \left( G_N(x,t)+\frac{\partial}{\partial \alpha}\right)
\cdot
{\det}_{N}(\hU_{jk}-\alpha\hQ_j\hQ_k)\ketd\right|_{\alpha=0},
\label{corr7}
\ea
where
\be
G_N(x,t) = \frac{1}{L} \sum_{n=-\infty}^{\infty} 
\frac{1}{2\pi\hat\rho(\lambda_n)} e^{\psi(\lambda_n)+\tau(\lambda_n)}, 
\ee
and

\ba\label{hU}
\hU_{jk} & = &
\frac{1}{L}\sum_{n=-\infty}^\infty
\frac{{\dis e^{\psi(\lambda_n)+\tau(\lambda_n)}
e^{-\frac12(\psi(\mu_j)+\psi(\mu_k)+\tau(\mu_j)+\tau(\mu_k))}
}}{2\pi\hat\rho(\lambda_n)}         
\non
&& 
\times 
\left\{t(\mu_k,\lambda_n) e^{-\phi_{D_1}(\lambda_n)}
-t(\lambda_n,\mu_k) e^{-\phi_{A_2}(\lambda_n)}
\right\}\non
&&\times 
\left\{t(\mu_j,\lambda_n) e^{-\phi_{D_1}(\lambda_n)}
-t(\lambda_n,\mu_j) e^{-\phi_{A_2}(\lambda_n)}
\right\},\label{hV}
\ea
\ba\label{hQ}
\hQ_j&=&
\frac{1}{L}\sum_{n=-\infty}^\infty
\frac{{\dis e^{\psi(\lambda_n)+\tau(\lambda_n)}
e^{-\frac12(\psi(\mu_j)+\tau(\mu_j))}}}{2\pi\hat\rho(\lambda_n)} 
\non
&&\times  \left\{t(\mu_j,\lambda_n) 
e^{-\phi_{D_1}(\lambda_n)}  
-t(\lambda_n,\mu_j) e^{-\phi_{A_2}(\lambda_n)}
\right\}.
\ea

Formula \Eq{corr7} is the determinant representation for the quantum
correlation function in a finite volume.
\section{Thermodynamic limit}

In order to calculate the correlation function in the ground state one
should consider the  limit where the number of particles and the 
length of the box tend to infinity with fixed constant density:
$N\to\infty,\quad L\to\infty, \quad N/L=D=const$.  In this limit 
the parameters $\{\lambda_n\}$ are described by distribution density
$\hat\rho(\lambda)$
$$
\hat\rho(\lambda)=\frac{1}{2\pi}\left(1-\frac{i}{L}\phi'_1(\lambda)
\right).
$$
(see Appendix A).
The sums in the expressions for $\hU_{jk}$ and $\hQ_j$ can be replaced
by corresponding integrals. Let us introduce the new function 
$Z(\lambda,\mm)$ 
\be\label{Z} 
Z(\lambda,\mm)=\frac{e^{-\phi_{D_1}(\lambda)}}{h(\mm,\lambda)}+
\frac{e^{-\phi_{A_2}(\lambda)}}{h(\lambda,\mm)}.
\ee
Then we can rewrite \Eq{hU} and \Eq{hQ} as
\ba\label{hU1}
\hU_{jk}& = &{\dis
\frac{1}{L}\sum_{n=-\infty}^\infty
\frac{(ic)^2e^{\psi(\lambda_n)+\tau(\lambda_n)}}{\hrp(\ll)
(\ll-\mm_j)(\ll-\mm_k)}}\non
&&{\dis\times
e^{-\frac12(\psi(\mm_j)+\psi(\mm_k)+\tau(\mu_j)+\tau(\mu_k))}
Z(\ll,\mm_k)Z(\ll,\mm_j)},
\ea
\be\label{hQ1}
\hQ(\mm) = 
\frac{1}{L}\sum_{n=-\infty}^\infty
\frac{ice^{\psi(\lambda_n)+\tau(\lambda_n)}
e^{-\frac12(\psi(\mm)+\tau(\mu))}}{\hrp(\ll)
(\ll-\mm)}Z(\ll,\mm).
\ee
Here $\hQ_j=-\hQ(\mm_j),~~\hQ_k=-\hQ(\mm_k)$. Using formula
\Eq{extrsing}, we get
\ba\label{hU2}&&\hskip3cm{\dis
\hU_{jk}
=L\delta_{jk}\frac{(ic)^2\hrp(\mm_j)}
{4\sin^2\frac{L}{2}\hxxj}Z^2(\mm_j,\mm_j)}\non
&&{\dis 
-\frac{(ic)^2}{2}\delta_{jk}\cot\frac L2\hxxj 
\frac{\partial}{\partial\mm_j} 
\left[e^{\frac12(\psi(\mm_j)-\psi(\mm_k)+\tau(\mm_j)-\tau(\mm_k))}
Z(\mm_j,\mm_k)Z(\mm_j,\mm_j)\right.}\non
&&\hskip3cm{\dis 
-\left.e^{-\frac12(\psi(\mm_j)-\psi(\mm_k)+\tau(\mm_j)-\tau(\mm_k))} 
Z(\mm_k,\mm_j)Z(\mm_k,\mm_k)
\right]_{\mm_k=\mm_j}}\non
&&{\dis 
-\frac{(ic)^2}{2}\frac{1-\delta_{jk}}{\mm_j-\mm_k}
\left[e^{\frac12(\psi(\mm_j)-\psi(\mm_k)+\tau(\mm_j)-\tau(\mm_k))}
Z(\mm_j,\mm_k)Z(\mm_j,\mm_j)\cot\frac L2\hxxj\right.}\non
&&\hskip3cm{\dis 
-\left.e^{-\frac12(\psi(\mm_j)-\psi(\mm_k)+\tau(\mm_j)-\tau(\mm_k))} 
Z(\mm_k,\mm_j)Z(\mm_k,\mm_k)\cot\frac L2\hxxk  
\right]}\non
&&\hskip3cm{\dis 
+\frac{(ic)^2}{2\pi(\mm_j-\mm_k)}
\pint \,d\lambda\left(\frac1{\lambda-\mm_j}- \frac1{\lambda-\mm_k}\right)
e^{\psi(\lambda)+\tau(\lambda)}}\non
&&\hskip3cm{\dis \times
e^{-\frac{1}{2}(\psi(\mm_j)+\tau(\mm_j)+\psi(\mm_k)+\tau(\mm_k))}
Z(\lambda,\mm_k)Z(\lambda,\mm_j)+\co.}
\ea
Here we denote  the principal value  by the symbol
$$
\pint\frac{d\lambda(\cdot)}{\lambda-\mu}\equiv
\mbox{V.P.}\stint\frac{d\lambda(\cdot)}{\lambda-\mu}=
\frac12\stint\frac{d\lambda(\cdot)}{\lambda-\mu+i0}+
\frac12\stint\frac{d\lambda(\cdot)}{\lambda-\mu-i0}
$$

Using \Eq{result} we calculate the sum \Eq{hQ1}:
\ba\label{hQ2}
\hQ(\mm)&=&{\dis \frac{ic}{2\pi}
\pint \frac{d\lambda}{\lambda-\mm}\hPPo Z(\lambda,\mm)}\non
&&{\dis 
-\frac{ic}2e^{\frac12(\psi(\mm)+\tau(\mm))}
Z(\mm,\mm)\cot\frac L2\hxx+\co.}
\ea

Function $\hxx$ is defined in Appendix A as
$$
\hxx=\mm-\frac{i}{L}\phi_1(\mm),
$$
(see \Eq{defxi}), and hence
$$
\begin{array}{rcl}{\dis 
\cot\frac{L}{2}\hxx}
&=&
{\dis 
i\frac{e^{iL\mm+\phi_1(\mm)}+1}
{e^{iL\mm+\phi_1(\mm)}-1},}\\
&&\vspace{-0.5em}\\
{\dis 
\sin^2\frac{L}{2}\hxx}
&=&
{\dis 
\frac{1}{4}\left(2-
e^{iL\mm+\phi_1(\mm)}-e^{-iL\mm-\phi_1(\mm)}\right).}
\end{array}
$$

Let us turn back to the formula \Eq{corr7}. We can move all 
$e^{p_1(\mm_m)}$ to the right vacuum $\ketd$. Then each operator 
$\phi_1$ entering into $\hU$ and $\hQ$ should be 
replaced by the rule  (see \Eq{mean3})
$$ 
\prod_{m=1}^{N}e^{p_1(\mm_m)}e^{\phi_1(\mm_j)}=
\prod_{m=1}^{N}\frac{h(\mm_m,\mm_j)}{h(\mm_j,\mm_m)}
e^{\phi_1(\mm_j)} \prod_{m=1}^{N}e^{p_1(\mm_m)}.
$$
Taking into account Bethe equations \Eq{BETHEeq}
$$
e^{iL\mm_j} \prod_{m=1}^{N}\frac{h(\mm_m,\mm_j)}{h(\mm_j,\mm_m)}
=-1,\qquad(\mbox{$N$ --- even}),
$$
we get
$$
\begin{array}{rcl}{\dis
\prod_{m=1}^{N}e^{p_1(\mm_m)}\cot\frac{L}{2}\hxxj}
&=&
{\dis
i\frac{\omega_-(\mm_j)}{\omega_+(\mm_j)}
\prod_{m=1}^{N}e^{p_1(\mm_m)},}\\
\vspace{-0.1em}&\vspace{-0.1em}&\vspace{-0.1em}\\
{\dis
\prod_{m=1}^{N}e^{p_1(\mm_m)}\sin^2\frac{L}{2}\hxxj}
&=&
{\dis 
\left(\frac{\omega_+(\mm_j)}{2}\right)^2
\prod_{m=1}^{N}e^{p_1(\mm_m)},}
\end{array}
$$
where
\be\label{omega}
\omega_\pm(\mm)=e^{\phi_1(\mm)/2}\pm e^{-\phi_1(\mm)/2}.
\ee

Operator $\hrp(\lambda)$ also contains $\phi_1(\lambda)$, so it does not
commute with $p_1(\mu)$:
\be\label{dressedrho}
\prod_{m=1}^{N}e^{p_1(\mm_m)}\hrp(\mm_j)=
2\pi \hat R(\mm_j)\prod_{m=1}^{N}e^{p_1(\mm_m)}.
\ee
where
$$
2\pi \hat R(\mm)=
\left(1+\frac1L\sum_{m=1}^{N}K(\mm,\mm_m)-
\frac iL\phi_1'(\mm)\right).
$$

Hence we have  the new representation for the correlation function
\ba\label{corr8}&&{\dis
\langle \psi(0,0)\psi^\dagger(x,t)\rangle_N=
\frac{e^{-iht}c^{-2N}}{
{\det}_N \frac{\partial\varphi_j}{\partial\mu_k}}}\non
&&{\dis\times 
\brad\prod_{m=1}^{N}e^{p_{0}(\mm_m)}
 \left( G_N(x,t)+\frac{\partial}{\partial 
\alpha}\right)}\non
&&{\dis 
\left.\times {\det}_{N}\Bigl(
\widetilde U_{jk}-
\alpha\widetilde Q(\mm_j)\widetilde Q(\mm_k)+\co\Bigr)
\ketd\right|_{\alpha=0},}
\ea
where
\ba\label{wtU}&&\hskip3cm{\dis
\widetilde U_{jk}
=L\delta_{jk}\frac{(ic)^2 2\pi\hat 
R(\mm_j)}{\omega_+^2(\mm_j)}Z^2(\mm_j,\mm_j)} \non 
&&{\dis 
-i\frac{(ic)^2}{2}\delta_{jk}\frac{\omega_-(\mm_j)}{\omega_+(\mm_j)} 
\frac{\partial}{\partial\mm_j} 
\left[e^{\frac12(\psi(\mm_j)-\psi(\mm_k)+\tau(\mm_j)-\tau(\mm_k))}
 Z(\mm_j,\mm_k)Z(\mm_j,\mm_j)\right.}\non
&&\hskip3cm{\dis 
-\left.e^{-\frac12(\psi(\mm_j)-\psi(\mm_k)+\tau(\mm_j)-\tau(\mm_k))} 
Z(\mm_k,\mm_j)Z(\mm_k,\mm_k)
\right]_{\mm_k=\mm_j}}\non
&&{\dis 
-i\frac{(ic)^2}{2}\frac{1-\delta_{jk}}{\mm_j-\mm_k}
\left[e^{\frac12(\psi(\mm_j)-\psi(\mm_k)+\tau(\mm_j)-\tau(\mm_k))}
Z(\mm_j,\mm_k)Z(\mm_j,\mm_j)\frac{\omega_-(\mm_j)}{\omega_+(\mm_j)}  
\right.}\non
&&\hskip3cm{\dis 
-\left.e^{-\frac12(\psi(\mm_j)-\psi(\mm_k)+\tau(\mm_j)-\tau(\mm_k))} 
Z(\mm_k,\mm_j)Z(\mm_k,\mm_k)\frac{\omega_-(\mm_k)}{\omega_+(\mm_k)}  
\right]}\non
&&\hskip3cm{\dis 
+\frac{(ic)^2}{2\pi(\mm_j-\mm_k)}
\pint \,d\lambda\left(\frac1{\lambda-\mm_j}- \frac1{\lambda-\mm_k}\right)
e^{\psi(\lambda)+\tau(\lambda)}}\non
&&\hskip3cm{\dis \times
e^{-\frac{1}{2}(\psi(\mm_j)+\tau(\mm_j)+\psi(\mm_k)+\tau(\mm_k))}
Z(\lambda,\mm_k)Z(\lambda,\mm_j).}
\ea
\ba\label{wtQ}
\widetilde Q(\mm)&=&{\dis \frac{ic}{2\pi}
\pint \frac{d\lambda}{\lambda-\mm}\hPPo Z(\lambda,\mm)}\non
&&{\dis 
+\frac{c}2e^{\frac12(\psi(\mm)+\tau(\mm))}
Z(\mm,\mm)\frac{\omega_-(\mm)}{\omega_+(\mm)}.}
\ea

Let us simplify the formul\ae\ \Eq{wtU} and \Eq{wtQ}. First, in \Eq{wtU} the
term proportional to $1-\delta_{jk}$ is defined only for $j\ne k$. Let us 
continue this term for all $j$ and $k$ using the l'H\^{o}pital's rule for
$j=k$. Then
\ba\label{wtU1}&&\hskip3cm{\dis
\widetilde U_{jk}
=L\delta_{jk}\frac{(ic)^2 
2\pi\rho_L(\mm_j)}{\omega_+^2(\mm_j)}Z^2(\mm_j,\mm_j)} \non 
&&{\dis 
-i\frac{(ic)^2}{2}\frac{1}{\mm_j-\mm_k}
\left[e^{\frac12(\psi(\mm_j)-\psi(\mm_k)+\tau(\mm_j)-\tau(\mm_k))}
Z(\mm_j,\mm_k)Z(\mm_j,\mm_j)\frac{\omega_-(\mm_j)}{\omega_+(\mm_j)}  
\right.}\non
&&\hskip3cm{\dis 
-\left.e^{-\frac12(\psi(\mm_j)-\psi(\mm_k)+\tau(\mm_j)-\tau(\mm_k))} 
Z(\mm_k,\mm_j)Z(\mm_k,\mm_k)\frac{\omega_-(\mm_k)}{\omega_+(\mm_k)}  
\right]}\non
&&\hskip3cm{\dis 
+\frac{(ic)^2}{2\pi(\mm_j-\mm_k)}
\pint \,d\lambda\left(\frac1{\lambda-\mm_j}- \frac1{\lambda-\mm_k}\right)
e^{\psi(\lambda)+\tau(\lambda)}}\non
&&\hskip3cm{\dis \times
e^{-\frac{1}{2}(\psi(\mm_j)+\tau(\mm_j)+\psi(\mm_k)+\tau(\mm_k))}
Z(\lambda,\mm_k)Z(\lambda,\mm_j),}
\ea
where
\be\label{rhoL}
2\pi\rho_L(\mm)=
\left(1+\frac1L\sum_{m=1}^{N}K(\mm,\mm_m)\right).
\ee
Using the Sokhodsky formula
$$
\mbox{V.P.}\frac{1}{x}=\frac{1}{x\pm i0}\pm i\pi\delta(x),
$$
one can rewrite the  expressions \Eq{wtU1} and \Eq{wtQ} as follows
\ba\label{wtU2}&&{\dis
\widetilde U_{jk}
=L\delta_{jk}\frac{(ic)^2 2\pi\rho_L(\mm_j)}
{\omega_+^2(\mm_j)}Z^2(\mm_j,\mm_j)}
\non
&&{\dis 
+\frac{(ic)^2}{2\pi(\mm_j-\mm_k)}
\stint \,\frac{d\lambda}{\omega_+(\lambda)}
\left(\frac{e^{\frac{1}{2}\phi_1(\lambda)}}{\lambda-\mm_j+i0}+
\frac{e^{-\frac{1}{2}\phi_1(\lambda)}}{\lambda-\mm_j-i0}-
\frac{e^{\frac{1}{2}\phi_1(\lambda)}}{\lambda-\mm_k+i0}- 
\frac{e^{-\frac{1}{2}\phi_1(\lambda)}}{\lambda-\mm_k-i0}\right)
}\non 
&&\hskip3cm{\dis \times
e^{\psi(\lambda)+\tau(\lambda)}
e^{-\frac{1}{2}(\psi(\mm_j)+\tau(\mm_j)+\psi(\mm_k)+\tau(\mm_k))}
Z(\lambda,\mm_k)Z(\lambda,\mm_j),}
\ea
\be\label{wtQ2}
\widetilde Q(\mm)={\dis \frac{ic}{2\pi}
\stint \,\frac{d\lambda}{\omega_+(\lambda)}
\left(\frac{e^{\frac{1}{2}\phi_1(\lambda)}}{\lambda-\mm+i0}+ 
\frac{e^{-\frac{1}{2}\phi_1(\lambda)}}{\lambda-\mm-i0}\right)
\hPPo Z(\lambda,\mm).}
\ee

Now let us move the term proportional to the length
of the box $L$ out of determinant:
\ba\label{deter}
&{\dis
{\det}_N(\widetilde U_{jk}-\alpha\widetilde Q(\mm_j)
\widetilde Q(\mm_k))
=\prod_{a=1}^{N}\left[(ic)^2 2\pi\rho_L(\mm_a) L
\left(\frac{Z(\mm_a,\mm_a)}{\omega_+(\mm_a)}\right)^2\right]}&\non
\vspace{-0.5em}&\vspace{-0.5em}&\vspace{-0.5em}\non
&{\dis\times{\det}_N\left(
\frac{\widetilde U_{jk}-\alpha\widetilde Q(\mm_j)\widetilde Q(\mm_k)}
{(ic)^2 2\pi\rho(\mm_k) L}\cdot
\frac{\omega_+(\mm_j)\omega_+(\mm_k)}
{Z(\mm_j,\mm_j)Z(\mm_k,\mm_k)}\right).}&
\ea
One can move the product 
$\prod_{a=1}^{N}\left(\frac{Z(\mm_a,\mm_a)}{\omega_+(\mm_a)}\right)^2$ 
to the left vacuum $\brad$:
\ba\label{subtheorem}
&&{\dis
\brad\prod_{a=1}^{N}e^{p_0(\mm_a)}\left(\frac{Z(\mm_a,\mm_a)}
{\omega_+(\mm_a)}\right)^2=
\brad\prod_{a=1}^{N}e^{p_0(\mm_a)}\left(\frac{
e^{-p_{A_1}(\mm_a)}+e^{-p_{D_2}(\mm_a)}}
{e^{p_2(\mm_a)/2}+e^{-p_2(\mm_a)/2}}\right)^2}\non
&&{\dis \hskip5cm
\equiv\brad\prod_{a=1}^{N}{\cal P}(\mm_a).}
\ea
Now let us  move the product in the r.h.s. 
of \Eq{subtheorem} to the right vacuum. 
The only  operator in the expressions for $\widetilde U$ and  $\widetilde Q$ 
which does not commute with ${\cal P}(\mm_a)$ is
 $\psi(\lambda)=\phi_{0}(\lambda)+\phi_{A_1}(\lambda)+
\phi_{D_2}(\lambda)+\phi_{2}(\lambda)$. In order to move
$\prod\limits_{a=1}^{N}{\cal P}(\mm_a)$ through the determinant we 
use the following lemma.

\begin{lem}
For arbitrary  $M=1,2,\dots$  and arbitrary 
complex numbers  $\lambda_1\dots,\lambda_M$, 
$\beta_1\dots,\beta_M,$:

\be\label{theorem4}
{\cal P}(\mm_a)\prod_{m=1}^{M}e^{\beta_m\psi(\lambda_m)}\ketd=
\prod_{m=1}^{M}e^{\beta_m\psi(\lambda_m)}\ketd
\ee
\end{lem}
{\it Proof.}
~~The proof is straightforward:
$$
\begin{array}{c}{\dis
{\cal P}(\mm_a)\prod_{m=1}^{M}e^{\beta_m\psi(\lambda_m)}\ketd=
e^{p_0(\mm_a)}\left(\frac{
e^{-p_{A_1}(\mm_a)}+e^{-p_{D_2}(\mm_a)}}
{e^{p_2(\mm_a)/2}+e^{-p_2(\mm_a)/2}}\right)^2
\prod_{m=1}^{M}e^{\beta_m\psi(\lambda_m)}\ketd}\\
\vspace{-0.5em}\\
{\dis
=\prod\limits_{m=1}^{M}e^{\beta_m\psi(\lambda_m)}\ketd
\prod\limits_{m=1}^{M}[h(\mm_a,\lambda_m)
h(\lambda_m,\mm_a)]^{\beta_m}}\\
\vspace{-0.5em}\\
{\dis
\times\left(\frac{\prod\limits_{m=1}^{M}
[h(\lambda_m,\mm_a)]^{-\beta_m}+
\prod\limits_{m=1}^{M}[h(\mm_a,\lambda_m)]^{-\beta_m}}
{\prod\limits_{m=1}^{M}\left[\frac{h(\lambda_m,\mm_a)}
{h(\mm_a,\lambda_m)}\right]^{\beta_m/2}
+\prod\limits_{m=1}^{M}\left[\frac
{h(\mm_a,\lambda_m)}{h(\lambda_m,\mm_a)}\right]^{\beta_m/2}}\right)^2=
\prod\limits_{m=1}^{M}e^{\beta_m\psi(\lambda_m)}\ketd.}
\end{array}
$$
This proves the lemma.

Since the determinant in r.h.s. of \Eq{deter}, being a function of operator 
$\psi$, is some linear combination of products of the type
$\prod\limits_{m=1}^{M}e^{\beta_m\psi(\lambda_m)}$ (with different $M,
\{\beta\}$ and $\{\lambda\}$),  we can move $\prod_{a=1}^{N}
{\cal P}(\mm_a)$ to the right vacuum without changing the matrix elements 
of the determinant \Eq{deter}. Therefore we have
\ba\label{corr9}&{\dis
\langle \psi(0,0)\psi^\dagger(x,t)\rangle_N=
\frac{e^{-iht}}{{\det}_N 
\frac{\partial\varphi_j}{\partial\mu_k}}}&\non
&{\dis
\times\left.\brad\prod_{a=1}^{N}\Bigl(2\pi\rho_L(\mm_a) L\Bigr)
 \left( G_N(x,t)+\frac{\partial}{\partial 
\alpha}\right) \cdot {\det}_{N} 
\left(W_{jk}+\Co\right)\ketd\right|_{\alpha=0},}&
\ea
where
\be\label{kern}
W_{jk}= \delta_{jk}+\frac{1}{2\pi\rho_L(\mm_k) L}(V_{jk}-
\alpha P(\mm_j)P(\mm_k)),
\ee
and
\ba\label{mainkern}&&{\dis
V_{jk}=\frac{\omega_+(\mm_j)\omega_+(\mm_k)}
{2\pi(\mm_j-\mm_k)Z(\mm_j,\mm_j)Z(\mm_k,\mm_k)} }\non
&&{\dis
\times\stint \,\frac{d\lambda}{\omega_+(\lambda)}
\left(\frac{e^{\frac{1}{2}\phi_1(\lambda)}}{\lambda-\mm_j+i0}+
\frac{e^{-\frac{1}{2}\phi_1(\lambda)}}{\lambda-\mm_j-i0}-
\frac{e^{\frac{1}{2}\phi_1(\lambda)}}{\lambda-\mm_k+i0}- 
\frac{e^{-\frac{1}{2}\phi_1(\lambda)}}{\lambda-\mm_k-i0}\right)
}\non 
&&\hskip3cm{\dis \times
e^{\psi(\lambda)+\tau(\lambda)}
e^{-\frac{1}{2}(\psi(\mm_j)+\tau(\mm_j)+\psi(\mm_k)+\tau(\mm_k))}
Z(\lambda,\mm_k)Z(\lambda,\mm_j),}
\ea
\ba\label{proect}&&{\dis
P(\mm)=\frac{\omega_+(\mm)}{2\pi Z(\mm,\mm)}
\stint \,\frac{d\lambda}{\omega_+(\lambda)}
\left(\frac{e^{\frac{1}{2}\phi_1(\lambda)}}{\lambda-\mm+i0}+ 
\frac{e^{-\frac{1}{2}\phi_1(\lambda)}}{\lambda-\mm-i0}\right)}\non
&&{\dis\hskip3.5cm
\times\hPPo Z(\lambda,\mm).}
\ea

Recall that in the thermodynamic limit
\be\label{limitnorm}
{\det}_N\frac{\partial\varphi_j}{\partial\mm_k}
\to\prod_{a=1}^{N}\Bigl(2\pi\rho(\mm_a)L\Bigr)
\det(\hat I-\frac{1}{2\pi}\hat K),\qquad
\mbox{(see \Eq{normlimit})},
\ee
\be\label{Gxt}
G_N(x,t)\to G(x,t)=\frac{1}{2\pi}\int\limits_{-\infty}^{+\infty}
e^{\psi(\nu)+\tau(\nu)}\,d\nu,
\ee
$$
\rho_L(\lambda)\to\rho(\lambda),\qquad
\mbox{(see \Eq{density})},
$$

The formula \Eq{corr9} contains the 
 determinant of the matrix $W$. In the thermodynamic limit
it will turn into a determinant of an integral operator. The simplest way
to see this is to express $\det W$ in terms of traces of 
powers of the matrix
$(V-\alpha PP)$. The replacement of summation by integration (in the limit)
is straightforward and is explained in detail in  Section XI.4 of
\cite{K.B.I.}. Therefore the determinant tends to the Fredholm determinant.
Now we arrive at the main theorem.
\begin{thm}~~~In the thermodynamic limit,
the time-dependent correlation function have the following
Fredholm determinant formula.
\ba\label{corr10}{\dis
\langle \psi(0,0)\psi^\dagger(x,t)\rangle=
e^{-iht}\brad \left( G(x,t)+\frac{\partial}{\partial 
\alpha}\right)}
{\dis\times
\left.\frac{\det 
\left(\hat I+\frac1{2\pi}\hat V_{\alpha}\right)}{\det\left(\hat 
I-\frac{1}{2\pi}\hat K\right)}\ketd\right|_{\alpha=0}.}
\ea 
Here the integral operator $\hat V_{\alpha}$ is given by
\begin{eqnarray}
&&\left(\hat V_{\alpha}f\right)(\lambda)
=\int_{-q}^q\left(\hat V(\lambda, \mu)
-\alpha \hat P (\mu) \hat P (\lambda)\right)f(\mu) d\mu,
\end{eqnarray}
where  $q$ is  the value of spectral parameter on the Fermi surface.
Here the kernels $\hat V(\mm_1,\mm_2)$ and
$\hat P(\mu)$ are given by
\ba\label{mainkernel}&&{\dis
\hat V(\mm_1,\mm_2)=\frac{\omega_+(\mm_1)\omega_+(\mm_2)}
{2\pi(\mm_1-\mm_2)Z(\mm_1,\mm_1)Z(\mm_2,\mm_2)} }\non
&&{\dis
\times\stint \,\frac{d\lambda}{\omega_+(\lambda)}
\left(\frac{e^{\frac{1}{2}\phi_1(\lambda)}}{\lambda-\mm_1+i0}+
\frac{e^{-\frac{1}{2}\phi_1(\lambda)}}{\lambda-\mm_1-i0}-
\frac{e^{\frac{1}{2}\phi_1(\lambda)}}{\lambda-\mm_2+i0}- 
\frac{e^{-\frac{1}{2}\phi_1(\lambda)}}{\lambda-\mm_2-i0}\right)
}\non 
&&\hskip3cm{\dis \times
e^{\psi(\lambda)+\tau(\lambda)}
e^{-\frac{1}{2}(\psi(\mm_1)+\tau(\mm_1)+\psi(\mm_2)+\tau(\mm_2))}
Z(\lambda,\mm_2)Z(\lambda,\mm_1),}
\ea
and
\ba\label{proect1}&&{\dis
\hat P(\mm)=\frac{\omega_+(\mm)}{2\pi Z(\mm,\mm)}
\stint \,\frac{d\lambda}{\omega_+(\lambda)}
\left(\frac{e^{\frac{1}{2}\phi_1(\lambda)}}{\lambda-\mm+i0}+ 
\frac{e^{-\frac{1}{2}\phi_1(\lambda)}}{\lambda-\mm-i0}\right)}\non
&&{\dis\hskip3.5cm
\times\hPPo Z(\lambda,\mm),}
\ea
where
$$
\omega_+(\mm)=e^{\phi_1(\mm)/2}+e^{-\phi_1(\mm)/2},
$$
$$
Z(\lambda,\mm)=\frac{e^{-\phi_{D_1}(\lambda)}}{h(\mm,\lambda)}+
\frac{e^{-\phi_{A_2}(\lambda)}}{h(\lambda,\mm)},
$$
$$
\tau(\lambda)=it\lambda^2-ix\lambda.
$$
The integral operator $\hat{K}$ is given in $(2.24)$.
\end{thm}
{\bf  We want to emphasize that formula \Eq{corr10} is our main result.}

It is easy to show that it has the correct free fermionic limit.
If $c\to+\infty$  (free fermionic case) then all commutators
\Eq{commutators} of auxiliary ``momenta'' and ``coordinates'' go to zero
because in this limit $h(\lambda,\mu)\to 1$. Hence one can put all dual
fields $\phi_a(\lambda)=0$. In particular

$$
\psi(\lambda)=0,\qquad \phi_1(\lambda)=0,
$$
$$
\omega_+(\lambda)=2,\qquad Z(\lambda,\mm)=2.
$$
whereby we have
\be\label{freekernel}
\hat V(\mm_1,\mm_2)\stackrel{c\to\infty}{=}\frac{2}{\pi(\mu_1-\mu_2)}
\pint\,d\lambda\left(\frac{1}{\lambda-\mm_1}-\frac{1}{\lambda-\mm_2}
\right)e^{\tau(\lambda)-\frac{1}{2}\tau(\mm_1)-
\frac{1}{2}\tau(\mm_2)},
\ee
\be\label{freeproector}
\hat P(\mm)\stackrel{c\to\infty}{=}\frac{1}{\pi}
\pint\,d\lambda\frac{1}{\lambda-\mm}
e^{\tau(\lambda)-\frac{1}{2}\tau(\mm)}.
\ee
\be
\hat K\stackrel{c\to\infty}{=}0,
\ee
\be
G(x,t)\stackrel{c\to\infty}{=}\frac1{2\pi}\stint\,d\nu e^{\tau(\nu)}.
\ee
Substitution of these formul\ae~ into \Eq{corr10}
reproduces the result of \cite{K.S.1}. In order to obtain
Lenard's determinant formula \cite{Len} one should also put $t=0$.

\section*{Summary}

The main result of the paper is formula \Eq{corr10}. It represents
the correlation function of local fields (in the infinite volume) as a 
mean value of  a determinant of a 
Fredholm integral operator. In order to
obtain this formula we introduced an auxiliary Fock space and auxiliary
Bose fields (all of them belong to the same Abelian sub-algebra). This
is the first step in description of 
the correlation function. In  forthcoming
publications we shall describe the Fredholm determinant 
by a completely integrable
integro-differential equation. Then we shall solve this equation by means of
the Riemann-Hilbert problem and evaluate its long-distance asymptotic.

\section*{Acknowledgments}
We wish to thank Professor Y.~Matsuo for useful
discussions and A.~Waldron for  an assistance.
This work is partly supported by the National Science Foundation (NSF)
under Grant No. PHY-9321165, the Japan Society for
the Promotion of Science, the Russian Foundation of Basic Research under
Grant No. 96-01-00344 and INTAS-01-166-ext.

\appendix
\section{Summation of singular expressions}

Let us consider equation \Eq{mainequation} 
\be\label{eq}
L\lambda_n-i\phi_1(\ll)=2\pi n.
\ee
where $i\phi_1(\lambda)$ is a real and bounded function for
$\Im\lambda=0$. Let us introduce a function $\hxi(\lambda)$
\be\label{defxi}
\hxi(\lambda)=\lambda-\frac{i}{L}\phi_1(\lambda).
\ee
Obviously
\be\label{xin}
\hxi(\ll)=\frac{2\pi n}{L}.
\ee
Comparing with \Eq{hrho} we get
\be\label{defrho}
\hrp(\lambda)=1-\frac{i}{L}\phi'_1(\lambda)=\hxi'(\lambda).
\ee
It follows from equation \Eq{eq} that
$$
|\lambda_{n+1}-\ll|\le\frac{2}{L}(\pi+M),
$$
where
$$
M=\sup\limits_{-\infty<\lambda<\infty}|\phi_1(\lambda)|.
$$
Hence, $|\lambda_{n+1}-\ll|\to 0$ if $L\to\infty$ and we can make the
following estimate
\be\label{estimation}
\frac{1}{L(\lambda_{n+1}-\ll)}=\hr(\ll)+\Co.
\ee
Due to \Eq{positive} we have $\hrp(\lambda)>0$.

During study of thermodynamic limit  the following sums appeared 
\be\label{sum} 
S=\frac 
1L\sum_{n=-\infty}^{\infty}\frac{f(\ll)}{\hrp(\ll)(\lm)}.  
\ee 
Here $f(\lambda)$ is some smooth function, $\mm$ is some fixed 
point on the real axis.  We shall be interested in the 
asymptotic of this sum when $L$ goes to infinity.

Let us present \Eq{sum} as the sum of three summands
\ba\label{partition}&&{\dis
S=\frac 1{2\pi L}\left(\sum_{n=-\infty}^{N_1-1}\frac{f(\ll)}{\hr(\ll)(\lm)}
+\right.}\non 
&&{\dis\left.
+\sum_{n=N_1}^{N_2}\frac{f(\ll)}{\hr(\ll)(\lm)}+
\sum_{n=N_2+1}^{\infty}\frac{f(\ll)}{\hr(\ll)(\lm)}\right).}
\ea
Here $N_1$ and $N_2$ are integers  such that in the limit
$L\to\infty$, the following properties are valid
\be\label{interval}
0<\mm-\lls<\infty,\qquad
0<\llb-\mm<\infty.
\ee
Obviously the first and the third summands in \Eq{partition} have no  
singularities in the domain of  summation. The corresponding 
sums are  integral sums, for example

\begin{eqnarray} 
&& S_1=\frac 
1L\sum_{n=-\infty}^{N_1-1}\frac{f(\ll)}{\hrp(\ll)(\lm)}= 
\sum_{n=-\infty}^{N_1-1}\frac{f(\ll)(\lambda_{n+1}-\ll)}{2\pi(\lm)}+\Co
\non 
&& 
\hskip5.5cm
=\frac{1}{2\pi}\int\limits_{-\infty}^\lls\frac{f(\lambda)}
{\lambda-\mm}\,d\lambda+\co. \label{leftsum}
\ea
An analogous formula is valid for $S_3$
\be\label{rightsum}
S_3=\frac 1L\sum_{n=N_2+1}^{\infty}\frac{f(\ll)}{\hrp(\ll)(\lm)}=
\frac{1}{2\pi}\int\limits_{\llb}^\infty\frac{f(\lambda)}
{\lambda-\mm}\,d\lambda+\co.
\ee

Consider the second summand in \Eq{partition}
$$
S_2=\frac1L\sum_{n=N_1}^{N_2}\frac{f(\ll)}{\hrp(\ll)(\lm)}.
$$
One can present $S_2$ in the following form
$$
S_2=S_2^{(1)}+S_2^{(2)},
$$
where
\ba\label{s22}&&{\dis
S_2^{(1)}=\frac1L\sum_{n=N_1}^{N_2}\left(\frac{f(\ll)}{\hrp(\ll)(\lm)}-
\frac{f(\mm)}{\hxi(\ll)-\hxx}\right),}\non
&&{\dis
S_2^{(2)}=\frac{f(\mm)}L\sum_{n=N_1}^{N_2}\frac1{\hxi(\ll)-\hxx}. }
\ea
Due to \Eq{defrho} $S_2^{(1)}$ has no singularities in the domain of
 summation. Therefore it can be replaced by the corresponding integral
 \be\label{int21}
S_2^{(1)}=\frac{1}{2\pi}\int\limits_\lls^\llb
\left(\frac{f(\lambda)}{\lambda-\mm}-
\frac{\hrp(\lambda) f(\mm)}{\hxi(\lambda)-\hxx}\right)
 \,d\lambda+\co.
 \ee
 Using \Eq{xin} we can rewrite $S_2^{(2)}$ in the following form
 $$
 S_2^{(2)}= \frac{f(\mm)}{2\pi}\sum_{n=N_1}^{N_2}\frac1
 {n-\frac L{2\pi}\hxx}.
 $$
 The last sum can be calculated explicitly in terms of the logarithmic 
 derivative of the $\Gamma$-function:
 $$
 \psi(x)=\frac d{dx}\ln\Gamma(x).
 $$
 We shall use the following properties of  the $\psi$-function 
 \ba\label{psiplus}
 &&{\dis \psi(x)+\frac 1x=\psi(x+1),}\\
 && \label{psiminus} {\dis 
 \psi(x)-\psi(1-x)=-\pi\cot\pi x,}\\
 &&\label{psiasimp} {\dis 
 \psi(x)\to\ln x+{\cal O}(1/x),\qquad x\to+\infty.}
 \ea
Now using \Eq{psiplus} we can write
\be\label{s22prim}
 S_2^{(2)}= \frac{f(\mm)}{2\pi}\left[\psi(N_2-\frac L{2\pi}\hxx+1)
-\psi(N_1-\frac L{2\pi}\hxx)\right].
\ee

The argument of the second $\psi$-function in \Eq{s22prim} is negative. 
Using \Eq{psiminus} one can flip the sign of this argument 
\begin{eqnarray}
 && {\dis S_2^{(2)}}={\dis \frac{f(\mm)}{2\pi}\left[
\psi(N_2-\frac L{2\pi}\hxx+1)\right.}\non
 &&
{\dis\left.-\psi(\frac L{2\pi}\hxx-N_1+1)
-\pi\cot\left(\frac L2\hxx\right)\right].} \label{s22second} 
\ea

Remember now that $0<\mm-\lls<\infty$ and $0<\llb-\mm<\infty$ (see 
\Eq{interval}). This means that the arguments of $\psi$-functions in 
\Eq{s22second} tend to infinity if $L\to\infty$. Therefore  we can use the 
asymptotic formula \Eq{psiasimp}
 \be\label{s21int} S_2^{(2)}= 
 \frac{f(\mm)}{2\pi}\left[\ln \left(\frac{N_2-\frac L{2\pi}\hxx} 
{\frac L{2\pi}\hxx-N_1}\right)-\pi\cot\left(\frac L2\hxx\right)\right]
+\co.  
\ee

Now let us turn back to \Eq{int21}. Let us present the r.h.s. as the
difference of two integrals. Both of them  should be understood in
the sense of principal value (V.P.):
 \be\label{difference}
S_2^{(1)}=\frac{1}{2\pi}\mint\frac{f(\lambda)}{\lambda-\mm}
\,d\lambda-\mint
\frac{f(\mm)\hr(\lambda)}{\hxi(\lambda)-\hxx}
 \,d\lambda+\co.
 \ee
Due to \Eq{defrho} one can compute the second term in \Eq{difference}
 explicitly
\ba\label{secondterm}{\dis
\mint\,d\lambda\frac{ 
 f(\mm)\hr(\lambda)}{\hxi(\lambda)-\hxx} = 
 \frac{f(\mm)}{2\pi}\mint\,\frac{d\hxi(\lambda)}{\hxi(\lambda)-\hxx}=}
\non
{\dis
 \frac{f(\mm)}{2\pi}\ln
 \left(\frac{\hxi(\llb)-\hxx}{\hxx-\hxi(\lls)}\right)=
\frac{f(\mm)}{2\pi}\ln\left(\frac{N_2-\frac L{2\pi}\hxx}{\frac 
 L{2\pi}\hxx-N_1}\right).}
\ea
Combining now \Eq{s21int}, \Eq{difference} and \Eq{secondterm} we get
$$
S_2=\frac{1}{2\pi}\mint\frac{f(\lambda)}{\lambda-\mm}
\,d\lambda- \frac{f(\mm)}2\cot\frac L2\hxx+\co.
$$
Finally, using \Eq{leftsum} and \Eq{rightsum} we find
\ba\label{result}&&{\dis
S=\frac 1L\sum_{n=-\infty}^{\infty}\frac{f(\ll)}{\hrp(\ll)(\lm)}}\non
\quad&&{\dis=
\frac{1}{2\pi}\pint\frac{f(\lambda)}{\lambda-\mm}
\,d\lambda
 - \frac{f(\mm)}2\cot\frac L2\hxx+\co.}
\ea
 This formula describes the asymptotic behavior of the sum \Eq{sum}.

For the evaluation of the thermodynamic limit 
of our determinant representation it is necessary to
consider  a  sum containing the second order pole
\be\label{second}
S'=\frac 1L\sum_{n=-\infty}^{\infty}\frac{f(\ll)}{\hrp(\ll)(\lm)^2}.
\ee
Taking the derivative of \Eq{result} with respect to $\mu$, we
get
\ba\label{result2}
S'&=&{\dis 
L\frac{\hrp(\mm)f(\mm)}{4\sin^2\frac L2\hxx}}\non
&+&{\dis
\frac{1}{2\pi}\frac\partial{\partial\mm}\pint\frac{f(\lambda)}
{\lambda-\mm}\,d\lambda-
\frac{1}{2}f'(\mm)\cot\frac L2\hxx+\co}.
\ea

We can use formul\ae\ \Eq{result} and \Eq{result2} to calculate
thermodynamic limit of \Eq{hU1}
\ba\label{twopoles}&{\dis
\frac{1}{L}\sum_{n=-\infty}^{\infty}
\frac{f(\ll|\mm_j,\mm_k)}{\hrp(\ll)(\ll-\mm_j)(\ll-\mm_k)}}&\non
&{\dis
=-\frac{1}{2(\mm_j-\mm_k)}\left[f(\mm_j|\mm_j,\mm_k)\cot\frac 
L2\hxi(\mm_j) -f(\mm_k|\mm_j,\mm_k)\cot\frac 
L2\hxi(\mm_k)\right]}\non 
&{\dis
+\frac 1{2\pi(\mu_j-\mu_k)}
\pint\,d\lambda f(\lambda|\mm_j,\mm_k)\left(\frac1{\lambda-\mm_j}
-\frac1{\lambda-\mm_k}\right)+\co.}&
\ea 
One should understand the r.h.s. 
of this equality by l'H\^{o}pital's rule if $j=k$. It is also useful to 
extract explicitly the  term proportional to the length of 
the box $L$ 
\ba\label{extrsing}
{}&{\dis 
\frac{1}{L}\sum_{n=-\infty}^{\infty}
\frac{f(\ll|\mm_j,\mm_k)}{\hrp(\ll)(\ll-\mm_j)(\ll-\mm_k)}=
\delta_{jk}L\frac{\hrp(\mm_j)f(\mm_j|\mm_j,\mm_j)}
{4\sin^2\frac L2\hxi(\mm_j)}}&\non
{}&{\dis
+\frac 1{2\pi(\mu_j-\mu_k)}
\pint\,d\lambda f(\lambda|\mm_j,\mm_k)\left(\frac1{\lambda-\mm_j}
-\frac1{\lambda-\mm_k}\right)}&\non
{}&{\dis
-\frac{1-\delta_{jk}}{2(\mm_j-\mm_k)}
\left[f(\mm_j|\mm_j,\mm_k)\cot\frac L2\hxi(\mm_j)
-f(\mm_k|\mm_j,\mm_k)\cot\frac L2\hxi(\mm_k)\right]}&\non
{}&{\dis
-\frac{\delta_{jk}}{2}\cot\frac L2\hxi(\mm_j)
\frac{\partial}{\partial\mm_j}\Biggl[f(\mm_j|\mm_j,\mm_k)
-f(\mm_k|\mm_j,\mm_k)\Biggr]_{\mm_j=\mm_k}+\co.}&
\ea
We have used formul\ae~\Eq{result} and \Eq{extrsing} in Section 6.

\section{Representation of dual fields}
 
What is the relation between our dual fields and the canonical Bose
fields. Canonical Bose fields $\psi_l(\lambda)$ can be characterized as follows
\be\label{commutpsi}
[\psi_l(\lambda),\psi_m^\dagger(\lambda)]=
\delta_{lm}\delta(\lambda-\mu),
\ee
(do not confuse this $\psi_l(\lambda)$ with the dual field
$\psi(\lambda)$) and
\be
\psi_l(\lambda)\ketd=0,\qquad
\brad\psi_l^\dagger(\lambda)=0.
\ee
The dual fields which appeared in this paper have the form
\be
\phi_a(\lambda)=q_a(\lambda)+p_a(\lambda),
\ee
where $p_a(\lambda)$ is the annihilation part
of $\phi_a(\lambda)$ and $q_a(\lambda)$ is its creation part.
Their commutation relations are
\be
[p_a(\lambda),q_b(\mu)]=\alpha_{ab}(\lambda,\mu).
\ee
Here $\alpha_{ab}(\lambda,\mu)$ is some complex function.
\be
p_a(\lambda)\ketd=0,\qquad \brad q_a(\lambda)=0
\ee
One can represent our $p_a$ and $q_b$ in terms of $\psi_l$ and 
$\psi_l^\dagger$, for example as 
\ba
p_a(\lambda)&=&\psi_a(\lambda),\non
\one&\one&\one\\
 q_b(\mu)&=&\sum_c\stint\,d\nu\alpha_{cb}(\nu,\mu)\psi_c^\dagger(\nu).
\ea 
 This shows that the dual fields, which appear in this paper are
linear combinations of the standard Bose fields. 

Let us now consider a related issue. We can realize the dual fields 
$\phi_1(\lambda)$ and $\phi_2(\lambda)$ as

$$ 
q_2(\lambda)=\psi_1^\dagger(\lambda),\qquad
p_1(\lambda)=\psi_1(\lambda);
$$
$$
q_1(\lambda)=\int\limits_{-\infty}^\infty \ln
\frac{h(\nu,\lambda)}{h(\lambda,\nu)}\psi_2^\dagger(\nu)\,d\nu,\qquad
p_2(\lambda)=\int\limits_{-\infty}^\infty \ln
\frac{h(\nu,\lambda)}{h(\lambda,\nu)}\psi_2(\nu)\,d\nu,
$$
Here $\dagger$ means Hermitian conjugation, and  
$$ 
[\psi_1(\lambda),\psi_2^\dagger(\mu)]=[\psi_2(\lambda),\psi_1^\dagger(\mu)]
=\delta(\lambda-\mu).
$$
Other commutators are equal to zero. These commutation relations differ 
from \Eq{commutpsi} only by  a trivial relabeling. Then
$$
\phi_2(\lambda)=\psi_1^\dagger(\lambda)+\psi_1(\lambda),\qquad
\phi_1(\lambda)=\int\limits_{-\infty}^\infty \ln
\frac{h(\nu,\lambda)}{h(\lambda,\nu)}(\psi_2^\dagger(\nu)
+\psi_2(\nu))\,d\nu.
$$
This means that $\phi_2(\lambda)$ and $i\phi_1(\lambda)$ are Hermitian 
operators:  
\be\label{hermit}
\phi_2^\dagger(\lambda)=\phi_2(\lambda),\qquad
(i\phi_1(\lambda))^\dagger=i\phi_1(\lambda),\qquad 
\mbox{for $\mathop{\rm Im}\lambda=0$}
\ee
After diagonalization they will turn into real functions.

\section{Reduction of number of dual fields}
We would like to reduce the number of dual fields in the determinant formula
for the correlation function in \Eq{corr10}. Here we shall show that
\be\label{reduce1}
 \phi_1(\lambda)=\phi_{D_1}(\lambda)-\phi_{A_2}(\lambda)
\qquad\mbox{and}\qquad \phi_2(\lambda)=0.
\ee
Recall the definition of the dual quantum fields \Eq{defdualfields} and
\Eq{commutators}
\be\label{defdualfields1}
 \begin{array}{lcl}{\dis
\phi_0(\lambda)=q_0(\lambda)+p_0(\lambda);}&&\\
{\dis
\phi_{A_j}(\lambda)=q_{A_j}(\lambda)+p_{D_j}(\lambda);}
&\qquad&{\dis
\phi_{D_j}(\lambda)=q_{D_j}(\lambda)+p_{A_j}(\lambda);}\\
{\dis
\phi_1(\lambda)=q_1(\lambda)+p_2(\lambda);}
&\qquad &{\dis
\phi_2(\lambda)=q_2(\lambda)+p_1(\lambda).}
\end{array}
\ee

\be\label{commutators1}\left\{
\begin{array}{lcl}
{\dis
[p_0(\lambda),q_0(\mu)]=\ln(h(\lambda,\mu)h(\mu,\lambda));}
&&\\
{\dis
[p_{D_j}(\lambda),q_{D_k}(\mu)]=\delta_{jk}\ln h(\lambda,\mu);}
&\qquad&
{\dis
[p_{A_j}(\lambda),q_{A_k}(\mu)]=\delta_{jk}\ln 
h(\mu,\lambda);}\\
{\dis
[p_1(\lambda),q_1(\mu)]
 = \ln\frac{h(\lambda,\mu)}{h(\mu,\lambda)};}
&\qquad& {\dis
[p_2(\lambda),q_2(\mu)]
=\ln\frac{h(\mu,\lambda)}{h(\lambda,\mu)}.}
\end{array} \right.
\ee

Remember also the definition of the field $\psi(\lambda)$ \Eq{psi}
$$
\psi(\lambda)=\phi_{0}(\lambda)+\phi_{A_1}(\lambda)
+\phi_{D_2}(\lambda)+\phi_{2}(\lambda).
$$
Notice that $q_1(\lambda)$ and
$p_2(\lambda)$ entering into $\phi_1(\lambda)$  do not commute only 
with $\psi(\mu)$:  
$$ 
\begin{array}{rcl} 
 [p_2(\lambda),\psi(\mu)]&=&\ln
{\dis\frac{h(\mu,\lambda)}{h(\lambda,\mu)},}\\
\vspace{-0.1em}&\vspace{-0.1em}& \vspace{-0.1em}\\
{} [\psi(\mu),q_1(\lambda)]&=&\ln
{\dis \frac{h(\mu,\lambda)}{h(\lambda,\mu)}.}
\end{array}
$$
On the other hand $q_{D_1}(\lambda)-q_{A_2}(\lambda)$ and
$p_{A_1}(\lambda)-p_{D_2}(\lambda)$ entering into
$\phi_{D_1}(\lambda)-\phi_{A_2}(\lambda)$ also
do not commute only with $\psi(\mu)$:  
$$ 
\begin{array}{rcl} 
[(p_{A_1}(\lambda)-p_{D_2}(\lambda)),\psi(\mu)]&=&\ln
{\dis\frac{h(\mu,\lambda)}{h(\lambda,\mu)},}\\
\vspace{-0.1em}&\vspace{-0.1em}& \vspace{-0.1em}\\
{}[\psi(\mu),(q_{D_1}(\lambda)-q_{A_2}(\lambda))]&=&\ln
{\dis \frac{h(\mu,\lambda)}{h(\lambda,\mu)},}
\end{array}
$$
so we can identify 
\be\label{phi1}
\phi_1(\lambda)=\phi_{D_1}(\lambda)-\phi_{A_2}(\lambda).
\ee
Then we can also put $\phi_2(\lambda)=q_2(\lambda)+p_1(\lambda)=0$
 because after the replacement
\Eq{phi1}, operators $q_2(\lambda)$ and $p_1(\lambda)$ commute with
everything.

Such a replacement implies 
\be\label{omegazet}
\omega_+(\lambda)=e^{\frac{1}{2}(\phi_{D_1}(\lambda)
+\phi_{A_2}(\lambda))}Z(\lambda,\lambda).
\ee
 
It means, that  Fredholm determinant in \Eq{corr10} realy depends
on three dual fields $\psi(\lambda)$, $\phi_{A_2}(\lambda)$
and $\phi_{D_1}(\lambda)$. We shall use this fact in our next
publications.

\section{Thermodynamics}

The thermodynamics of the quantum nonlinear 
Schr\"odinger equation was described
by C.~N.~Yang and C.~P.~Yang \cite{YY}. It involves
few equations. The central equation is for an energy of excitation
$\varepsilon(\lambda)$:
\be
\varepsilon(\lambda)=\lambda^2-h-
\frac{T}{2\pi}\stint\frac{2c}{c^2
+(\lambda-\mu)^2}\ln\left(1+e^{-\frac
{\varepsilon(\mu)}T}\right)\,d\mu.
\ee
Other important functions are the local density (in momentum space) of
particles $\rho_p(\lambda)$ and the total local density $\rho_t(\lambda)$
(it includes particles and holes). They satisfy equations:
\be
2\pi\rho_t(\lambda)=1+
\stint\frac{2c}{c^2
+(\lambda-\mu)^2}
\rho_p(\mu)\,d\mu,
\ee
\be
\frac{\rho_p(\lambda)}{\rho_t(\lambda)}=
\left(1+e^{\frac{\varepsilon(\lambda)}T}\right)^{-1}\equiv
\vartheta(\lambda).
\ee
The global density $D=N/L$ can be represented as
\be
D=\stint\rho_p(\lambda)\,d\lambda.
\ee

In order to obtain the determinant representation of 
the temperature correlation function we can use the following representation
\be\label{temperature}
\langle\psi(0,0)\psi^\dagger(x,t)\rangle_T\equiv
\frac{\tr\left( e^{-\frac HT}\psi(0,0)\psi^\dagger(x,t)\right)
}
{\tr e^{-\frac HT}}=
\frac{
\langle\Omega_T|\psi(0,0)\psi^\dagger(x,t)|\Omega_T\rangle}
{\langle\Omega_T|\Omega_T\rangle}.
\ee
Here $|\Omega_T\rangle$ is one of eigenfunctions of the Hamiltonian,
which is present in the state of thermo-equilibrium. It is proven
in Section I.8 of \cite{K.B.I.} that the r.h.s. of \Eq{temperature}
does not depend on the particular choice of
$|\Omega_T\rangle$.

Now we have to recalculate thermodynamic limit. 
First we shall return to the determinant representation of 
the correlation function in a finite volume, see
formul\ae~\Eq{corr9}--\Eq{proect}. We should also notice that the 
thermodynamic limit of square of the norm should be changed (comparing to
\Eq{limitnorm}) as follows:
\be\label{limitnormT}
{\det}_N\frac{\partial\varphi_j}{\partial\mm_k}
\to\prod_{a=1}^{N}\Bigl(2\pi\rho_t(\mm_a)L\Bigr)
\det(\hat I-\frac{1}{2\pi}\hat K_T),
\ee
see Section X.4 of \cite{K.B.I.}. Here $\mu_j$ 
correspond to $|\Omega_T\rangle$.
In the thermodynamic limit summation with 
respect to indicis $j$ and $k$ in
\Eq{corr9} will be replaced by integration $\rho_p(\mu_k)\,d\mu_k$.
Also $\rho_L(\mu)$ defined in \Eq{rhoL} goes to $\rho_t$:
$$
\rho_L(\mu)\to\rho_t(\mu).
$$
After dividing $\rho_p(\mu_k)\,d\mu_k$
by $\rho_t(\mu_k)$ which appears in the denominator \Eq{kern} we shall 
obtain an integration $\stint\vartheta(\lambda)\,d\lambda(\cdot)$ insted of
$\qint\,d\lambda(\cdot)$. The  details of these calculations
are explained in Section XI.5 of \cite{K.B.I.}.

The integral operator $\hat K_T$ can be defined by its kernel
\be\label{kernT}
K_T(\mu_1,\mu_2)=\left(\frac{2c}{c^2+(\mu_1-\mu_2)^2}\right)
\sqrt{\vartheta(\mu_1)} 
\sqrt{\vartheta(\mu_2)}.
\ee

Now let us formulate the final formula for the 
representation of the temperature
correlation function of local fields in the thermodynamic limit 
\ba{\dis
\langle \psi(0,0)\psi^\dagger(x,t)\rangle=
e^{-iht}\brad \left( G(x,t)+\frac{\partial}{\partial
\alpha}\right)}
{\dis\times
\left.\frac{\det
\left(\hat I+\frac1{2\pi}\hat V_{\alpha,T}\right)}{\det\left(\hat
I-\frac{1}{2\pi}\hat K\right)}\ketd\right|_{\alpha=0}.}
\ea
Here the integral operator $\hat V_{\alpha,T}$ is given by
\begin{eqnarray}
\left(\hat V_{\alpha,T}f\right)(\lambda)
=\int_{-\infty}^\infty\left(\hat V_{T}(\lambda, \mu)
-\alpha \hat P_{T} (\mu) \hat P_{T} (\lambda)\right)f(\mu) d\mu.
\end{eqnarray}
Here the kernel of $\frac1{2\pi}\biggl(\hat V_T(\mu_1,\mu_2)-\alpha \hat 
P_T(\mm_1)\hat P_T(\mm_2)\biggr)$ differs from the zero temperature case
by the measure and limits of integration: 
\be\label{kernelT}
\hat V_T(\mu_1,\mu_2)-\alpha \hat 
P_T(\mm_1)\hat P_T(\mm_2)=
\biggl(\hat V(\mu_1,\mu_2)-\alpha \hat 
P(\mm_1)\hat P(\mm_2)\biggr)
\sqrt{\vartheta(\mu_1)}
\sqrt{\vartheta(\mu_2)}.
\ee
It acts on the whole real axis $-\infty<\mu<\infty$. Here 
$\hat V(\mu_1,\mu_2)$
is given exactly by \Eq{mainkernel}, $\hat P(\mu)$ is given by formula
\Eq{proect1} and $G(x,t)$ is given by \Eq{Gxt}.

\end{document}